\documentclass[aps,prd,preprint,tightenlines,nofootinbib,showpacs,superscriptaddress]{revtex4-1}
\usepackage[colorlinks=true]{hyperref}
\hypersetup{colorlinks=true, citecolor=blue, urlcolor=blue, linkcolor=blue}
\usepackage{graphicx}
\usepackage{epsfig}
\usepackage{epstopdf}
\usepackage{subfigure}
\usepackage{amsmath}
\usepackage{cancel}
\usepackage{amssymb}
\usepackage{hyperref}
\usepackage{rotating}   
\usepackage{xcolor}      
\usepackage{ulem}

\begin{document}
\arraycolsep1.5pt
\newcommand{\Ima}{\textrm{Im}}
\newcommand{\Rea}{\textrm{Re}}
\newcommand{\mev}{\textrm{ MeV}}
\newcommand{\gev}{\textrm{ GeV}}

\title{How to reveal the nature of three or more pentaquark states?}

\author{C. W. Xiao}
\affiliation{School of Physics and Electronics, Central South University, Changsha 410083, China}

\author{J. X. Lu}
\affiliation{School of Physics, Beihang University, Beijing 100191, China}

\author{J. J. Wu}
\affiliation{School of Physical Sciences, University of Chinese Academy of Sciences (UCAS), Beijing 100049, China}

\author{L. S. Geng}
\email{lisheng.geng@buaa.edu.cn}
\affiliation{School of Physics, Beihang University, Beijing 100191, China}
\affiliation{Beijing Advanced Innovation Center for Big Data-based Precision Medicine, Beihang University, Beijing 100191, China}
\affiliation{Beijing Key Laboratory of Advanced Nuclear Materials and Physics, Beihang University, Beijing 100191, China}
\affiliation{School of Physics and Microelectronics, Zhengzhou University, Zhengzhou, Henan 450001, China}

\date{\today}

\begin{abstract}

Within the chiral unitary approach and with the constraints of heavy quark spin symmetry, we study the coupled channel interactions of ${\bar D}^{(*)}\Sigma_c^{(*)}$ channels, close to whose thresholds three pentaquark-like $P_c$ states have been reported by the LHCb Collaboration. 
In the present work, we take into account the contributions of pion exchanges via box diagrams to the interaction potentials, and therefore lift the degeneracy in the masses of ${\bar D}^*\Sigma_c^{(*)}$ spin multiplets. 
Fitting the $J/\psi p$ invariant mass distributions in the $\Lambda_b^0 \to J/\psi K^- p$ decay, we find that the LHCb pentaquark states can not be reproduced in the direct $J/\psi p$ production in the $\Lambda_b^0$ decay, and can only be indirectly produced in the final state interactions of the $\Lambda_b^0$ decay products, ${\bar D}^*\Sigma_c^{(*)}$,  which
further supports the nature of these states as $\bar{D}\Sigma_c$ molecules. 
Based on the fit results obtained, we study the partial decay widths/branching ratios to other decay channels, $\bar{D}^* \Lambda_c$, $\bar{D} \Lambda_c$, and $\eta_c N$, and the corresponding invariant mass distributions. 
The resonances with $J^P=\frac{1}{2}^-$, $P_c(4312)$, $P_c(4440)$ and the one of $\bar{D}^* \Sigma_c^*$ around 4500 MeV, have large partial decay width into $\eta_c N$, and thus, can be easily seen in the $\eta_c N$ invariant mass distributions. 
By contrast, the states with $J^P=\frac{3}{2}^-$, $P_c(4457)$, the (predicted) narrow $P_c(4380)$ and the bound state of $\bar{D}^* \Sigma_c^*$ with a mass of about 4520 MeV, do not decay into $\eta_c N$. 
Therefore, the $\eta_c N$ channel should be studied in the future to provide further insights into the nature of these states, especially that of the $P_c(4440)$ and $P_c(4457)$. 

\end{abstract}
\pacs{}

\maketitle

\section{Introduction}

In 2015, two pentaquark-like resonances are reported by the LHCb Collaboration in the $J/\psi p$ mass spectrum of the $\Lambda_b^0 \to J/\psi K^- p$ decay~\cite{Aaij:2015tga}, referred to as $P_c(4380)^+$ and $P_c(4450)^+$, of which the masses and  widths are
\begin{align*}
&M_{P_{c1}} = (4380 \pm 8 \pm 29)\mev, \quad
  \Gamma_{P_{c1}}= (205 \pm 18 \pm 86)\mev,  \\
  &M_{P_{c2}} = (4449.8 \pm 1.7 \pm 2.5)\mev, \quad
  \Gamma_{P_{c2}}= (39 \pm 5 \pm 19)\mev,
\end{align*}
with some uncertainties about their spin-parity $J^P$ quantum numbers~\cite{Aaij:2015fea}. 
Later, these two $P_c$ states are confirmed by a model-independent re-analysis of the experimental data \cite{Aaij:2016phn}, and also observed in the $\Lambda_b^0 \to J/\psi p \pi^-$ decay \cite{Aaij:2016ymb} as suggested in Refs. \cite{Burns:2015dwa,Wang:2015pcn}. 
In fact, these pentaquark-like states with hidden charm have been predicted before the experimental findings in the early works \cite{Wu:2010jy,Wu:2010vk,Wang:2011rga,Yang:2011wz,Yuan:2012wz,Wu:2012md,Garcia-Recio:2013gaa,Xiao:2013yca,Uchino:2015uha,Karliner:2015ina} using different theoretic models. 
In Ref. \cite{Wu:2010vk}, it is suggested to search for these hidden charm molecular states in the decay channel of $J/\psi N$, which is later studied in more detail in Ref. \cite{Molina:2012mv}. 
The cross sections of the $J/\psi N$ and $\eta_c N$ channels are investigated to search for signals of these $P_c$ states in Ref.~\cite{Xiao:2015fia}, based on the interactions with their coupled channels. 
Indeed, the coupled channel effects are important for the dynamical productions of these pentaquark-like states~\cite{Wu:2010jy}, as concluded in Ref.~\cite{Skerbis:2018lew}, where the $P_c$ resonances are not observed in the lattice QCD study of single channel scattering of $J/\psi N$ and $\eta_c N$.  
After the discovery of the LHCb Collaboration, the multi-quark states have attracted renewed interests, which can be seen in the recent reviews \cite{Chen:2016qju,Hosaka:2016pey,Chen:2016spr,Lebed:2016hpi,Esposito:2016noz,Guo:2017jvc,Ali:2017jda,Olsen:2017bmm,Karliner:2017qhf,Yuan:2018inv,Liu:2019zoy,Brambilla:2019esw}. 
In 2019, the LHCb Collaboration updated the results of Ref.~\cite{Aaij:2015tga}, where three clear narrow structures are reported \cite{Aaij:2019vzc},
\begin{align*}
&M_{P_{c1}} = (4311.9\pm0.7^{+6.8}_{-0.6})\, {\rm MeV}, \quad
  \Gamma_{P_{c1}}= (9.8\pm2.7^{+3.7}_{-4.5})\, {\rm MeV},  \\
&M_{P_{c2}} = (4440.3\pm1.3^{+4.1}_{-4.7})\, {\rm MeV}, \quad
  \Gamma_{P_{c2}}= (20.6\pm4.9^{+8.7}_{-10.1})\, {\rm MeV}, \\
&M_{P_{c3}} = (4457.3\pm0.6^{+4.1}_{-1.7})\, {\rm MeV}, \quad
  \Gamma_{P_{c3}}= (6.4\pm2.0^{+5.7}_{-1.9})\, {\rm MeV}.
\end{align*}
From the updated results, one can see that the original peak of $P_c(4450)$ is now split into two states of $P_c(4440)$ and $P_c(4457)$, and a fluctuation observed in the original spectrum has given rise to a new narrow resonance $P_c(4312)$. 
Whereas, the broad $P_c(4380)$ can neither be confirmed nor refuted in the new spectrum \cite{Aaij:2015tga}, where some structures around this energy region can also be seen.
Even though, there are many theoretical supports from QCD sum rules for the $P_c(4380)$ resonance \cite{Chen:2015moa,Azizi:2016dhy,Azizi:2018bdv,Azizi:2017bgs,Azizi:2018dva,Ozdem:2018qeh}.

The new findings of three $P_c$ states have also attracted much theoretical and experimental interests. 
The $P_c(4312)$, $P_c(4440)$, and $P_c(4457)$ are often assumed to be molecular states of $\bar{D}^{(*)}\Sigma_c^{(*)}$ with $J^P=\frac{1}{2}^-$, $\bar{D}^*\Sigma_c$ with $J^P=\frac{1}{2}^-$ and $\bar{D}^*\Sigma_c$ with $J^P=\frac{3}{2}^-$, because of their closeness to
the thresholds of respective channels~ \cite{Chen:2019asm,Chen:2019bip,Liu:2019tjn,He:2019ify,Xiao:2019mst,Guo:2019kdc,Xiao:2019aya,Zhang:2019xtu,Wu:2019rog,Wang:2019ato,Zhang:2020erj,Xu:2020gjl,Peng:2020gwk}, but, there are some other assignments for the components of $\bar{D}^{(*)}\Sigma_c^{(*)}$ and the spin-parity quantum numbers~ \cite{Chen:2019bip,Liu:2019tjn,Xu:2020gjl,Peng:2020gwk,Shimizu:2019ptd,Zhu:2019iwm,Wang:2019got,Cheng:2019obk,Yamaguchi:2019seo,Liu:2019zvb,Pan:2019skd}. 
Note that, heavy quark spin symmetry (HQSS)~\cite{Neubert:1993mb,hqss00} predicts seven bound states in the single channel treatment of Ref. \cite{Liu:2019tjn}, of which some are consistent with the ones obtained in Ref. \cite{Xiao:2019aya} with the interactions also constrained by HQSS. 
In the compact diquark model~\cite{Ali:2019npk}, the $P_c(4312)$ is explained as an $S$-wave diquark-diquark-antiquark state with $J^P=\frac{3}{2}^-$, $P_c(4440)$ and $P_c(4457)$ as $P$-wave states with $J^P=\frac{3}{2}^+$ and $J^P=\frac{5}{2}^+$. 
Moreover, starting from the effective Lagrangians respecting chiral and heavy quark symmetry in the Bethe-Salpeter framework \cite{Xu:2020gjl}, two $P_c(4457)$ states are predicted with spin parities of $J^P=\frac{3}{2}^-$ and $J^P=\frac{1}{2}^-$ and nearly degenerate masses, and thus, there are four molecular states not only three. 
Similarly, in Ref. \cite{Peng:2020gwk} the likely existence of two peaks is proposed for the $P_c(4457)$ state with $J^P=\frac{1}{2}^\pm$, when the $\bar{D} \Lambda_c(2595)$ is taken into account for its close threshold as firstly introduced and studied in Refs. \cite{Geng:2017hxc,Burns:2019iih}. 
By contrast, using the $S$-matrix approach and performing a systematic analysis of the reaction amplitudes, the authors in Ref. \cite{Fernandez-Ramirez:2019koa} explained
the $P_c(4312)$ as a virtual state. 
The molecular picture for these $P_c$ states is contrasted with the hadrocharmonium picture in Ref.~\cite{Eides:2019tgv}. 
Ref.~\cite{Guo:2019fdo} suggested that the molecular nature of the $P_c(4457)$ resonance can be checked by studying its isospin breaking decay channel of $J/\psi \Delta$ in experiments. 
On the other hand, it is not so optimistic to reveal more features of these $P_c$ states in the present experimental results of the $P_c$ photoproduction in the $\gamma p \to J/\psi p$ process as discussed in Ref. \cite{Cao:2019kst}, which proposes that the $\bar{D} \Lambda_c$ channel would be essential for searching for these $P_c$ states in photoproduction. 
Using an effective Lagrangian approach, the photoproduction of these $P_c$ states is also investigated in Refs.~\cite{Wang:2019krd,Wu:2019adv,Cao:2019kst} and it is suggested that higher precision experimental data are needed.
Indeed, there is no evidence for the three $P_c$ resonances in the measurement of the $\gamma p \to J/\psi p$ cross section by the GlueX experiment \cite{Ali:2019lzf} with not enough statistics, where the molecular model can not be ruled out with the upper limits of the branching fractions of $P_c \to J/\psi p$. 
A further study about the photoproduction of these pentaquark states at RHIC and LHC can be found in Ref.~\cite{Goncalves:2019vvo}, and the electroproduction in Refs. \cite{Xie:2020niw, Yang:2020eye} at these and future EicC (Electron-ion collider in China) facilities. 
Lately, the D0 Collaboration reported their confirmatory evidence for these $P_c$ states with the data collected at the Fermilab Tevatron collider \cite{Abazov:2019kwn}. 
Furthermore, a different type of photoproduction reaction, $\gamma p \to \bar{D}^{*0} \Lambda_c^+$, is proposed in Ref. \cite{Huang:2016tcr} for finding the $P_c$ states, which does not have the kinematic effects of the triangle singularity as in the $\Lambda_b^0 \to J/\psi p \pi^-$ decay \cite{Guo:2015umn,Liu:2015fea,Mikhasenko:2015vca}. 
In addition, searching for these $P_c$ states in the  $\pi^- p \to J/\psi n$ reaction is suggested in Ref. \cite{Wang:2019dsi}, and, the reaction $\pi^- p \to D^- \Sigma_c^+$ is proposed in Ref. \cite{Garzon:2015zva} to look for the $\bar{D} \Sigma_c$ bound state. 

Based on the mass spectrum of these $P_c$ states, the work of \cite{Du:2019pij} claims the existence of a narrow $P_c(4380)$ in addition to the three $P_c$ states by fitting the $J/\psi p$ invariant mass distributions as commented in Ref. \cite{Xiao:2019aya}, and predicts three other molecular states as found in Refs. \cite{Liu:2019tjn,Xiao:2019aya,Yamaguchi:2019seo}. 
Analysing the $J/\psi p$ spectroscopy with the $K$-matrix method, Ref. \cite{Kuang:2020bnk} assigns the $P_c(4312)$ as a $\bar{D}\Sigma_c$ molecule, $P_c(4440)$ a $S$-wave compact pentaquark state and $P_c(4457)$ as a cusp effect. 
In the present work, based on the results of Ref.~\cite{Xiao:2019aya}, we study the $J/\psi p$ invariant mass distributions in the $\Lambda_b^0 \to J/\psi p \pi^-$ decay using the chiral unitary approach (ChUA) to describe the coupled channel interactions. 
More details about this approach can be found in the recent reviews \cite{Oller:2019opk,MartinezTorres:2020hus,Oller:2020guq,Guo:2020hli}. 
In the previous work of Ref.~\cite{Xiao:2019aya}, the two ${\bar D}^*\Sigma_c$ states, assigned as the $P_c(4440)$ and $P_c(4457)$, are degenerate. 
Thus, we first introduce the pion exchange potentials \cite{Uchino:2015uha} to split their masses to better describe the experimental data. 
Indeed, the pion exchange potentials introduced in the box diagrams are crucial for the degeneracy breaking of the $\Lambda_b(5912)$ and $\Lambda_b(5920)$ states in the $B^* N$ interactions \cite{Liang:2014eba}, which is extended to the interactions of $DN$ and $D^* N$ with their coupled channels in Ref. \cite{Liang:2014kra} for reproducing the two $\Lambda_c$ states, $\Lambda_c(2595)$ and $\Lambda_c(2625)$. 
In the following, we first introduce the ChUA briefly. 
Next, we show our fit results with $J/\psi p$ directly produced in the $\Lambda_b^0 \to J/\psi p \pi^-$ decay, and then, our results with $J/\psi p$ indirectly produced in the final state interactions. With the fit results obtained, we calculate the couplings to all the coupled channels, the partial decay widths (branching ratios), and predict the invariant mass distributions to the other possible decay channels for these $P_c$ states. 
Finally, we conclude with a short summary.

\section{Formalism}

Following Ref. \cite{Xiao:2013yca}, the Bethe-Salpeter equation is used for the coupled channel interactions in the isospin $I=1/2$ sector, with seven coupled channels of $\eta_c N$, $J/\psi N$, $\bar{D}\Lambda_c$, $\bar{D}\Sigma_c$, $\bar{D}^*\Lambda_c$, $\bar{D}^*\Sigma_c$, and $\bar{D}^*\Sigma_c^*$ for spin parity $J^P=1/2^-$, and five channels of $J/\psi N$, $\bar{D}^*\Lambda_c$, $\bar{D}^*\Sigma_c$, $\bar{D}\Sigma_c^*$, $\bar{D}^*\Sigma_c^*$ for $J^P=3/2^-$. 
In addition, there is a single channel of $\bar{D}^*\Sigma_c^*$ for $J^P=5/2^-$. 
More details about the interactions for other isospin sectors can be found in Ref. \cite{Xiao:2013yca}, where there is no bound state as expected due to the repulsive interaction potentials.  
The Bethe-Salpeter equation in matrix form is adopted for evaluating the scattering amplitudes,
\begin{equation}
T = [1 - V \, G]^{-1}\, V,
\label{eq:BS}
\end{equation}
where $G$ is the loop functions with  meson -baryon intermediate states and the potential $V$ respecting HQSS is given in Tables~\ref{tab:vij11} and \ref{tab:vij31} for the $J=1/2, \, I=1/2$ and $J=3/2, \, I=1/2$ sectors, respectively, where the coefficients $\mu_{i}^I$, $\mu_{ij}^I$ ($i,j=1,2,3$) and $\lambda_2^I$ are unknown low energy constants with the HQSS constraint. 
More details can be found in Ref. \cite{Xiao:2013yca}. 
Note that, in the sector of $J=5/2, \, I=1/2$, there is only one channel, $\bar{D}^*\Sigma_c^*$, for which the potential is attractive and generates a bound state~\cite{Xiao:2013yca}. 
Since this state can not be coupled to the $J/\psi N$ channel as discussed in Ref.~\cite{Xiao:2019aya}, we do not consider it in the present work, and we focus on the properties of the three $P_c$ states in the $J/\psi p$ invariant mass distributions.

\begin{table}
     \renewcommand{\arraystretch}{1.7}
     \setlength{\tabcolsep}{0.4cm}
\centering
\caption{Potential matrix elements  $V_{ij}$ of Eq.~(\ref{eq:BS}) in the  $J=1/2,~I=1/2$ sector.}
\label{tab:vij11}
\begin{tabular}{ccccccc}
\hline\hline
$\eta_c N$ & $J/\psi N$ &  $\bar D \Lambda_c$ &  $\bar D \Sigma_c$ &  $\bar D^* \Lambda_c$
  &  $\bar D^* \Sigma_c$ &  $\bar D^* \Sigma^*_c$   \\
\hline
$\mu_1$ & 0 & $\frac{\mu_{12}}{2}$ &
 $\frac{\mu_{13}}{2}$ & $\frac{\sqrt{3} \mu_{12}}{2}$ &
 $-\frac{\mu_{13}}{2 \sqrt{3}}$ & $\sqrt{\frac{2}{3}} \mu_{13}$ \\
  & $\mu_1$ & $\frac{\sqrt{3} \mu_{12}}{2}$ & $-\frac{\mu_{13}}{2 \sqrt{3}}$ & $-\frac{\mu_{12}}{2}$
 & $\frac{5 \mu_{13}}{6}$ & $\frac{\sqrt{2}\mu_{13}}{3}$ \\
  &  & $\mu_2$ & 0 & 0 & $\frac{\mu_{23}}{\sqrt{3}}$ & $\sqrt{\frac{2}{3}} \mu_{23}$ \\
  &  &  & $\frac{1}{3} (2 \lambda_2 + \mu_3)$ & $\frac{\mu_{23}}{\sqrt{3}}$ & $\frac{2 (\lambda_2 - \mu_3)}{3 \sqrt{3}}$ & $\frac{1}{3} \sqrt{\frac{2}{3}} (\mu_3-\lambda_2 )$ \\
  &  &  &  & $\mu_2$ & $-\frac{2 \mu_{23}}{3}$ & $\frac{\sqrt{2} \mu_{23}}{3}$ \\
  &  &  &  &  & $\frac{1}{9} (2 \lambda_2 +7 \mu_3)$ & $\frac{1}{9} \sqrt{2} (\mu_3-\lambda_2)$ \\
  &  &  &  &  &  & $\frac{1}{9} (\lambda_2+8 \mu_3)$ \\
\hline\hline
\end{tabular}
\end{table}

\begin{table}
     \renewcommand{\arraystretch}{1.7}
     \setlength{\tabcolsep}{0.4cm}
\centering
\caption{Potential matrix elements  $V_{ij}$ of Eq.~(\ref{eq:BS}) in the $J=3/2,~I=1/2$ sector.}
\label{tab:vij31}
\begin{tabular}{ccccc}
\hline\hline
$J/\psi N$ &  $\bar D^* \Lambda_c$ &  $\bar D^* \Sigma_c$
&  $\bar D \Sigma^*_c$  &  $\bar D^* \Sigma^*_c$   \\
\hline
$\mu_1$ & $\mu_{12}$ & $\frac{\mu_{13}}{3}$ & $-\frac{\mu_{13}}{\sqrt{3}}$ & $\frac{\sqrt{5} \mu_{13}}{3}$ \\
  & $\mu_2$ & $\frac{\mu_{23}}{3}$ & $-\frac{\mu_{23}}{\sqrt{3}}$ & $\frac{\sqrt{5} \mu_{23}}{3}$ \\
  &  & $\frac{1}{9} (8 \lambda_2 + \mu_3)$ & $\frac{\lambda_2 - \mu_3}{3 \sqrt{3}}$ & $\frac{1}{9} \sqrt{5} (\mu_3-\lambda_2)$  \\
  &  &  & $\frac{1}{3} (2 \lambda_2 +\mu_3)$ &
   $\frac{1}{3} \sqrt{\frac{5}{3}} (\lambda_2 -\mu_3)$  \\
  &  &  &  & $\frac{1}{9} (4 \lambda_2 +5 \mu_3)$  \\
\hline\hline
\end{tabular}
\end{table}

\begin{figure}
\centering
\includegraphics[width=0.48\linewidth]{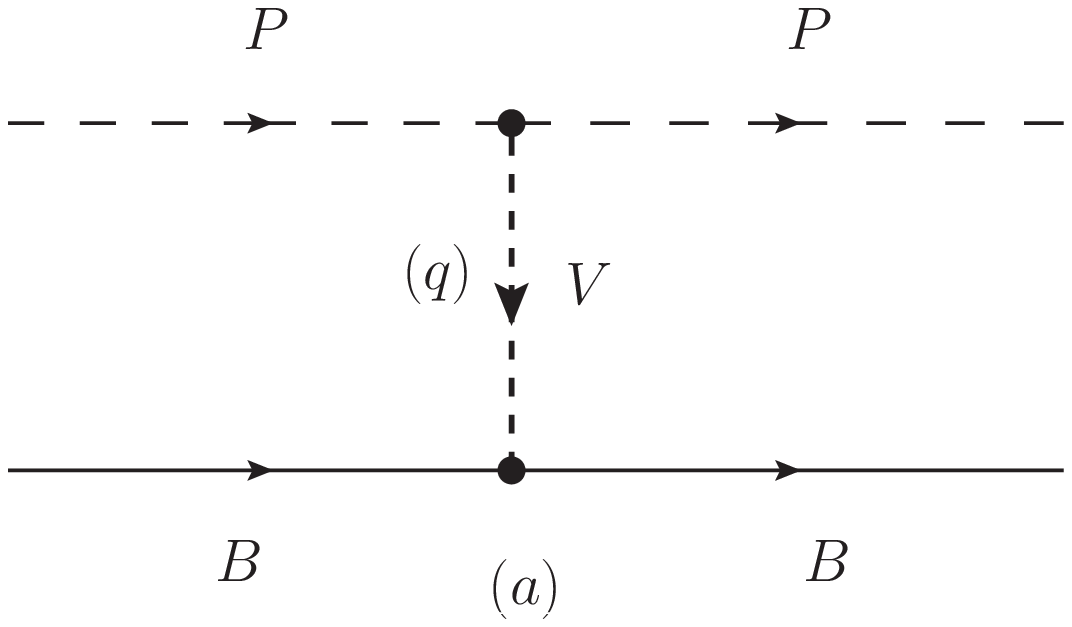}
\includegraphics[width=0.48\linewidth]{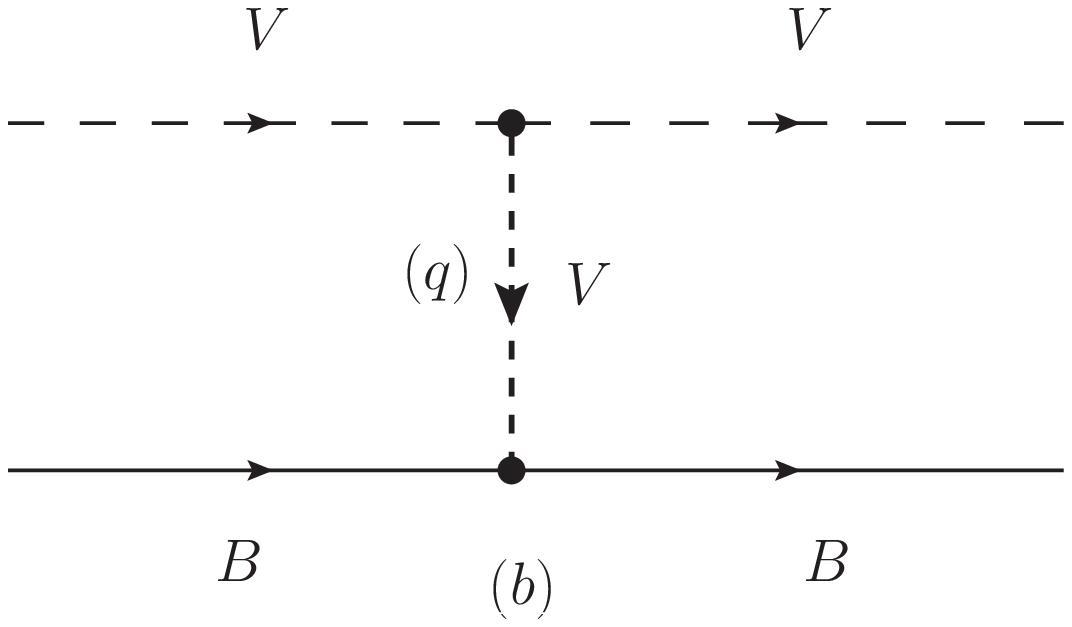}
\caption{Diagrams for the pseudoscalar-baryon (PB) interaction (a) and vector-baryon (VB) interaction (b) with the exchange of vector mesons.}
\label{fig:PVB}
\end{figure}

There are seven parameters under the HQSS constraint, which just depend on the isospin ($I$) and are independent of the spin $J$. 
In the present work, we take the same constraints as those in Ref. \cite{Xiao:2013yca}, which rely on the use of the extended local hidden gauge approach \cite{Bando:1984ej,Bando:1987br,Meissner:1987ge},  with the dynamics for the interactions originating from the exchange of vector mesons, as shown in Fig. \ref{fig:PVB}. 
These constraints for all the $I=1/2$ sectors are given by
\begin{equation}
\begin{split}
\mu_1 &= 0, \quad \mu_{23} = 0, \quad \lambda_2 = \mu_3, \quad \mu_{13} = -\mu_{12}, \\
\mu_2 &= \frac{1}{4f^2} (k^0 + k'^0),\quad \mu_3 = -\frac{1}{4f^2} (k^0 + k'^0),\\
\mu_{12} &= -\sqrt{6}\ \frac{m_\rho^2}{p^2_{D^*} - m^2_{D^*}}\ \frac{1}{4f^2}\ (k^0 + k'^0),\label{eq:ji11fi}
\end{split}
\end{equation}
where $f_\pi = 93 \mev$, $m_{D^*}$ is the $D^*$ mass, $k^0$ and $k'^0$ are the energies of the mesons in the $P(V)B \to P'(V') B'$ transition at tree level, and $p^2_{D^*} $ comes from the exchanged $D^*$ at tree level of some suppressed transitions (for example $\eta_c N\to \bar{D}\Lambda_c$), which are given by
\begin{align}
k^0=& \frac{s+m_M^2-m_B^2}{2\sqrt{s}}, \\
k'^0=& \frac{s+m_{M'}^2-m_{B'}^2}{2\sqrt{s}}, \\
p^2_{D^*}=&m_M^2+m_{M'}^2-2\,k^0\, k'^0,
\end{align}
where $m_M$ ($m_{M'}$) and $m_B$ ($m_{B'}$) are the masses of the incoming (outgoing) meson and baryon, respectively, and $s$ is the Mandelstam variable of the meson-baryon system.
Please note that the $\mu_{23}=0$ means that the $\bar{D}^*\Lambda$ channel is decoupled from the $\bar{D}^{(*)}\Sigma^{(*)}_c$ channels, which will lead to the fact that the $P_c$ states are almost entirely generated from $\bar{D}^{(*)}\Sigma^{(*)}_c$ channels and have very small partial decay width to the $\bar{D}^*\Lambda$ channel (see our results later).

In addition, the propagator matrix $G$ is a diagonal matrix with elements of meson-baryon loop functions. Using the dimensional regularization, they are given by
\footnote{A general expression for n-dimensions can be found in {\it e.g.} Ref. \cite{Djukanovic:2009gt}.}
\begin{equation}
\begin{split}
G_i (s) =& \frac{2 M_i}{16 \pi^2} \Big\{ a_\mu + \ln \frac{M_i^2}{\mu^2} + \frac{m_i^2 - M_i^2 +s}{2s} \ln \frac{m_i^2}{M_i^2}  \\
& + \frac{q_{cmi}}{\sqrt{s}} \big[ \ln(s - (M_i^2 - m_i^2) + 2 q_{cmi} \sqrt{s})  \\
& + \ln(s + (M_i^2 - m_i^2) + 2 q_{cmi} \sqrt{s})  \\
& - \ln(-s - (M_i^2 - m_i^2) + 2 q_{cmi} \sqrt{s})  \\
& - \ln(-s + (M_i^2 - m_i^2) + 2 q_{cmi} \sqrt{s}) \big] \Big\},  \\
\end{split}
\label{eq:Gdr}
\end{equation}
where $m_i$, $M_i$ are the masses of meson and baryon in the $i^{\rm th}$ channel, respectively, and $q_{cmi}$ is the three-momentum of the $i^{\rm th}$ channel in the center-of-mass (CM) frame, given by
\begin{equation}
q_{cmi} (s)=\frac{\lambda^{1/2}(s, M_i^2, m_i^2)}{2\sqrt{s}}\, ,
\end{equation}
with the usual K\"allen triangle function $\lambda(a,b,c)=a^2+b^2+c^2-2(ab+ac+bc)$. 
Therefore, the free parameters are $a_\mu$ and $\mu$.  
Note that they are not independent but correlated with each other, see the second term $\ln \frac{M_i^2}{\mu^2}$ in Eq. \eqref{eq:Gdr}, and more discussions can be found in Refs. \cite{Oller:2000fj,Ozpineci:2013zas}. 
Thus, in practice, we fix the value of $\mu$ as $\mu=1 \gev$ (the so called natural value \cite{Oller:2000fj}), and more discussions will be provided later.

\begin{figure}
\centering
\includegraphics[width=0.48\linewidth]{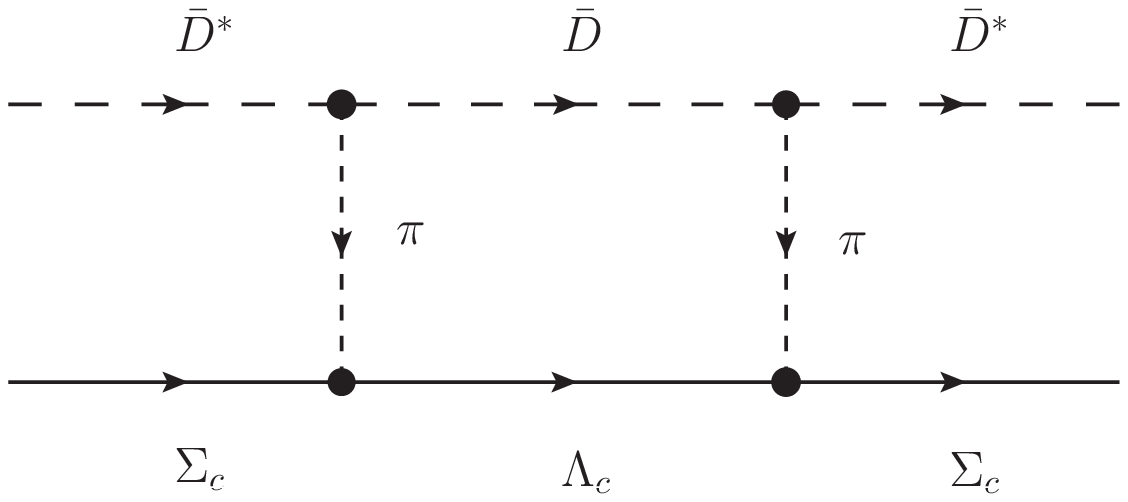}
\includegraphics[width=0.48\linewidth]{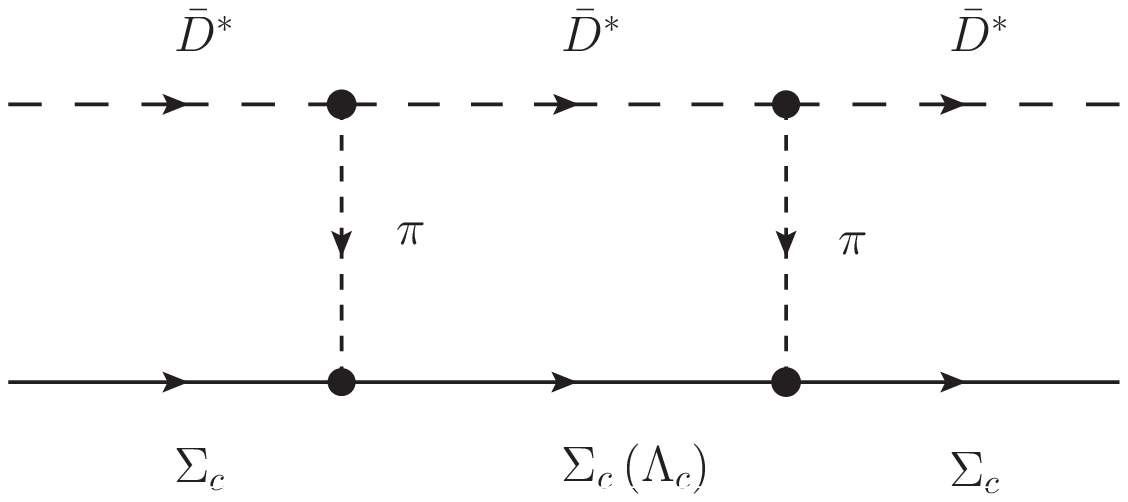}
\caption{Box diagram contributions with the intermediate state of $\bar{D}\Lambda_c$ (left) and $\bar{D}^*\Sigma_c\,(\Lambda_c)$ (right) for the ${\bar D}^*\Sigma_c$ channel in the sector of $J^P=\frac{1}{2}^-$.}
\label{fig:box}
\end{figure}

In Ref.~\cite{Xiao:2019aya}, the multiplets of ${\bar D}^*\Sigma_c$ and ${\bar D}^*\Sigma_c^*$ with different spin $J$ are nearly degenerate, where the two ${\bar D}^*\Sigma_c$ states are assigned as the $P_c(4440)$ and $P_c(4457)$, because of their different widths. 
Thus, as discussed in the introduction, to break the degeneracy, we will add the corrections of the pion exchange potentials via
box diagrams, as done in Ref. \cite{Uchino:2015uha}. 
In principle, $\pi$ exchange interaction could be included systematically between the channels as discussed in Refs. \cite{Xiao:2013yca,Voloshin:2019aut}, i.e., the off-diagonal potential, where, for example, such $\mu_{23}$ can not be zero as shown in Eqs. \eqref{eq:ji11fi}.
And correspondingly, although these box diagrams would be automatically included through such $\pi$ exchange interaction, these single $\pi$ exchange will make the calculation much more complicated \footnote{Though the pion exchange potential has been discussed in Ref. \cite{Xiao:2013yca}, where they found that the contributions from the pion exchange are small compared with the vector exchange potential, one should be careful with the singularities in the left hand cut when the pion exchange with large momentum transfer are taken into account, as pointed out in Ref. \cite{Gulmez:2016scm} in the case of $\rho\rho$ interactions and further discussed for the unphysical effects in Refs. \cite{Geng:2016pmf,Molina:2019rai}.}, which is left for our future work.
Here the box diagrams shown in Fig. \ref{fig:box} can be recognized as the first order correction of the potential of $\bar{D}^*\Sigma_c \to \bar{D}^*\Sigma_c$ process, and later it will be shown to be enough for the explanation of the mass splitting between $P_c(4440)$ and $P_c(4450)$.  
However, as found in Ref. \cite{Uchino:2015uha},  the pion exchange was not negligible and brought large corrections to binding energies to all the poles when the box diagram contributions were taken into account for all the coupled channels. 
The effect of the box diagram corrections is stronger in the charm sector \cite{Liang:2014kra,Uchino:2015uha} than the one in the beauty sector \cite{Liang:2014eba}. 
To not strongly distort the spectrum obtained in Ref.~\cite{Xiao:2019aya}, which already agrees
with the LHCb data reasonably well, we limit the corrections of the box diagram contributions only to the channels of ${\bar D}^*\Sigma_c$ and ${\bar D}^*\Sigma_c^*$ in the $J=\frac{1}{2}$ sector to break the mass degeneracy, since there are some structures in the region around the threshold of ${\bar D}^*\Sigma_c^*$ in the $J/\psi p$ invariant mass distributions \cite{Aaij:2019vzc} and three molecular states are predicted in this region too \cite{Liu:2019tjn,Xiao:2019aya,Du:2019pij}. 
In principle, one can also add the box diagram contributions to the $J=\frac{3}{2}$ sector, but, it is found to be difficult to assign the one with a larger width having $J=\frac{1}{2}$ as the $P_c(4457)$ in our results \footnote{We have studied this alternative and found that the corresponding fit is not good close to the $P_c(4440)$.}. 
Note that, the assignment, $J^P=\frac{1}{2}^-$ for $P_c(4457)$ and $J^P=\frac{3}{2}^-$ for $P_c(4440)$, is also not favoured in Ref. \cite{Wang:2019ato} where a systematic study is performed in the framework of the heavy hadron chiral perturbation theory. 
As found in Ref.~\cite{Uchino:2015uha}, for the $\bar{D}^* \Sigma_c$ channel, the box diagram contributions come from the channels of $\bar{D}\Lambda_c$, $\bar{D}^* \Sigma_c$ and $\bar{D}^* \Lambda_c$, as depicted in Fig. \ref{fig:box}. 
Indeed, the contributions from these $\bar{D}^{(*)}\Lambda_c$ channels are very important for the reproduction of these $P_c$ states as found in Ref. \cite{Wang:2019ato}, some of which are taken into account in Ref. \cite{Ke:2019bkf} as well.
We show the formalism for the box diagram contributions in detail below. 
Following Ref. \cite{Uchino:2015uha}, the normal box corrections from the $\bar{D}\Lambda_c$ channel (as shown on the left panel of Fig. \ref{fig:box}) and the $\bar{D} \Sigma_c^*$ contributions are,
\begin{align}
\delta V \left( \bar{D}^* \Sigma_c \to \bar{D} \Lambda_c \to \bar{D}^* \Sigma_c; J=1/2 \right) &=
  REL2 \times FAC \times \left(\frac{\partial I_1^{'}}{\partial m_{\pi}^2} + 2 I_2^{'} + I_3^{'} \right),  \\
\delta V \left( \bar{D}^* \Sigma_c^* \to \bar{D} \Sigma_c^* \to \bar{D}^* \Sigma_c^*; J=1/2 \right) &=
  REL3 \times FAC \times \frac{\partial I_1^{'}}{\partial m_{\pi}^2},
\end{align}
and the ones stemming from the anomalous term, see the right panel of Fig. \ref{fig:box}, are
\begin{align}
\delta V_{\rm an} \left( \bar{D}^* \Sigma_c \to \bar{D}^* \Sigma_c \to \bar{D}^* \Sigma_c \right)  &=
  REL1 \times AFAC \times \frac{\partial I_1^{'}}{\partial m_{\pi}^2},  \\
\delta V_{\rm an} \left( \bar{D}^* \Sigma_c \to \bar{D}^* \Lambda_c \to \bar{D}^* \Sigma_c \right)  &=
  REL2 \times AFAC \times \frac{\partial I_1^{'}}{\partial m_{\pi}^2} ,  \\
\delta V_{\rm an} \left( \bar{D}^* \Sigma_c^* \to \bar{D}^* \Sigma_c^* \to \bar{D}^* \Sigma_c^* \right) &=
  REL3 \times AFAC \times \frac{\partial I_1^{'}}{\partial m_{\pi}^2},
\end{align}
with the factors defined as
\begin{align}
FAC &= \frac{9}{2} g^2 \left( \frac{m_{D^*}}{m_{K^*}} \right)^2 \left( \frac{F+D}{2f} \right)^2 , \\
AFAC &=  \frac{9}{8} G^{'2} \left( \frac{D+F}{2f} \right)^2 m_{D^*}^2 ,\\
REL1 &= \frac{4}{9} \left( \frac{2F}{D+F} \right)^2,  \\
REL2 &=  \frac{1}{9} \left( \frac{2D}{D+F} \right)^2 , \\
REL3 &= \frac{5}{9} \left( \frac{f_{\Sigma^*}}{m_{\pi}} \right)^2/ \left( \frac{D+F}{2f} \right)^2= \frac{16}{45},
\end{align}
 where $D=0.75$ and $F=0.51$ ~\cite{Borasoy:1998pe} for the two couplings of the Yukawa vertex, $G' = \frac{3 m_V^2}{16 \pi^2 f^3_\pi}$ with $m_V\simeq 780 \mev$, $g=\frac{m_V}{2 f_\pi}$, and  the expressions of $I^{'}_1, I^{'}_2, I^{'}_3$ are given by~\cite{Liang:2014eba}
\begin{eqnarray}
I_1' &=& \int^{q_{max}} \frac{d^3 q}{(2\pi)^3} \frac{4}{3} \vec{q}\,^4 \frac{1}{2 \omega_B (\vec{q}\,)} \frac{M_N}{E_N (\vec{q}\,)} \frac{Num}{Den} F(\vec{q}\,) ,  \label{eq:box1}  \\
I_2' &=& \int^{q_{max}} \frac{d^3 q}{(2\pi)^3} 2 \vec{q}\,^2 \frac{1}{2 \omega_B (\vec{q}\,)} \frac{M_N}{E_N (\vec{q}\,)} \frac{Num}{Den} F(\vec{q}\,) ,  \\
I_3' &=& \int^{q_{max}} \frac{d^3 q}{(2\pi)^3} \frac{3}{2 \omega_B (\vec{q}\,)} \frac{M_N}{E_N (\vec{q}\,)} \frac{F(\vec{q}\,) }{P^0_{in} + K^0_{in} -E_N (\vec{q}\,) - \omega_B (\vec{q}\,) + i \epsilon},  \label{eq:box3}
\end{eqnarray}
with $q_{max}$ the cutoff, and
\begin{eqnarray}
Num &=& K^0_{in} - E_N (\vec{q}\,) - 2 \omega_\pi (\vec{q}\,) - \omega_B (\vec{q}\,) + P^0_{in},  \\
Den &=& 2 \omega_\pi (\vec{q}\,) [P^0_{in} - \omega_\pi (\vec{q}\,) - \omega_B (\vec{q}\,) + i \epsilon] [K^0_{in} - E_N (\vec{q}\,) - \omega_\pi (\vec{q}\,) + i \epsilon] \nonumber \\
 && \times [P^0_{in} + K^0_{in} - E_N (\vec{q}\,) - \omega_B (\vec{q}\,) + i \epsilon], \\
 F(\vec{q}\,) &=& \Big( \frac{\Lambda^2}{\Lambda^2 + \vec{q}\,^2} \Big)^2,
\end{eqnarray}
where $P^0_{in}$, $K^0_{in}$  are the energies for the incoming meson and baryon, respectively, and $F(\vec{q}\,)$ is the monopole form factor introduced in the Yukawa vertex, and $\Lambda \simeq 1 \gev$. 
For more details see Ref.~\cite{Liang:2014eba}. 
At the end, we add these box corrections to the potential of the corresponding channel, written as
\begin{equation}
V_{ij}=V_{ij} + \delta V + \cdots + \delta V_{\rm an} + \cdots,
\end{equation}
where the dots mean that there may be more than one box diagram contribution.

\section{Results with $J/\psi p$ produced directly}

\begin{figure}
\centering
\includegraphics[scale=0.7]{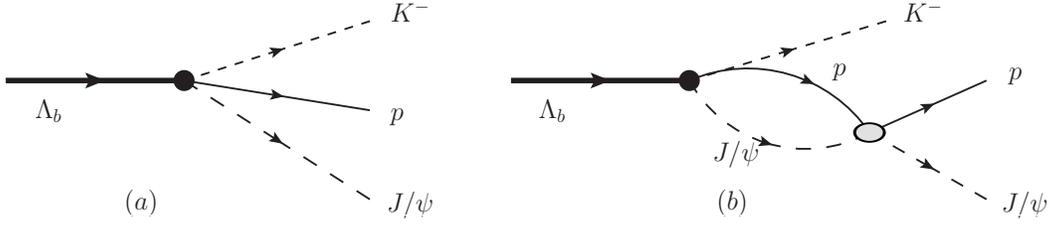}
\caption{Diagrams for the $\Lambda_b^0 \to J/\psi K^-\, p$ decay: (a) direct $J/\psi K^- \, p$ decay at tree level; (b) final state interactions of $J/\psi p$.}
\label{fig:lambdec}
\end{figure}

\begin{figure}
\centering
\includegraphics[width=0.48\linewidth]{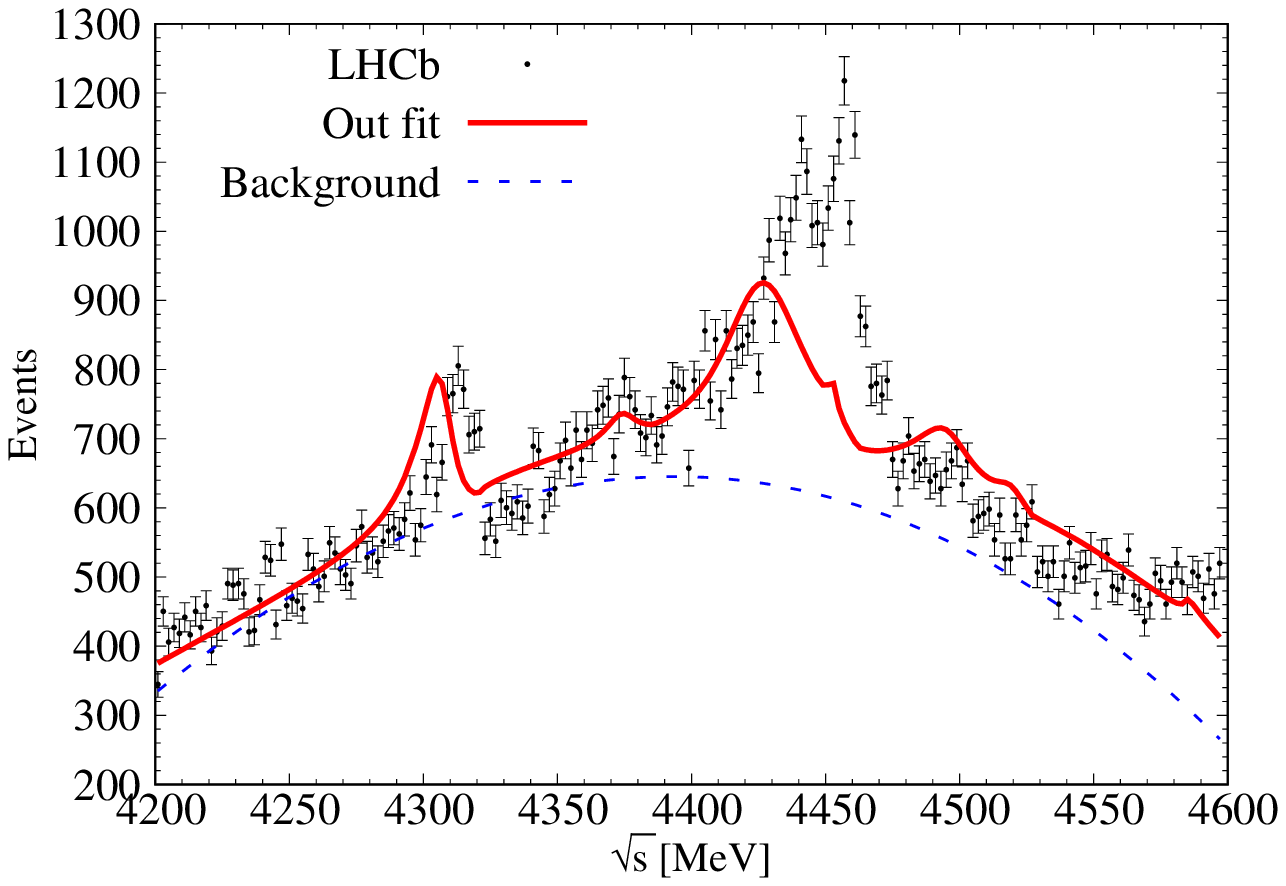}
\includegraphics[width=0.48\linewidth]{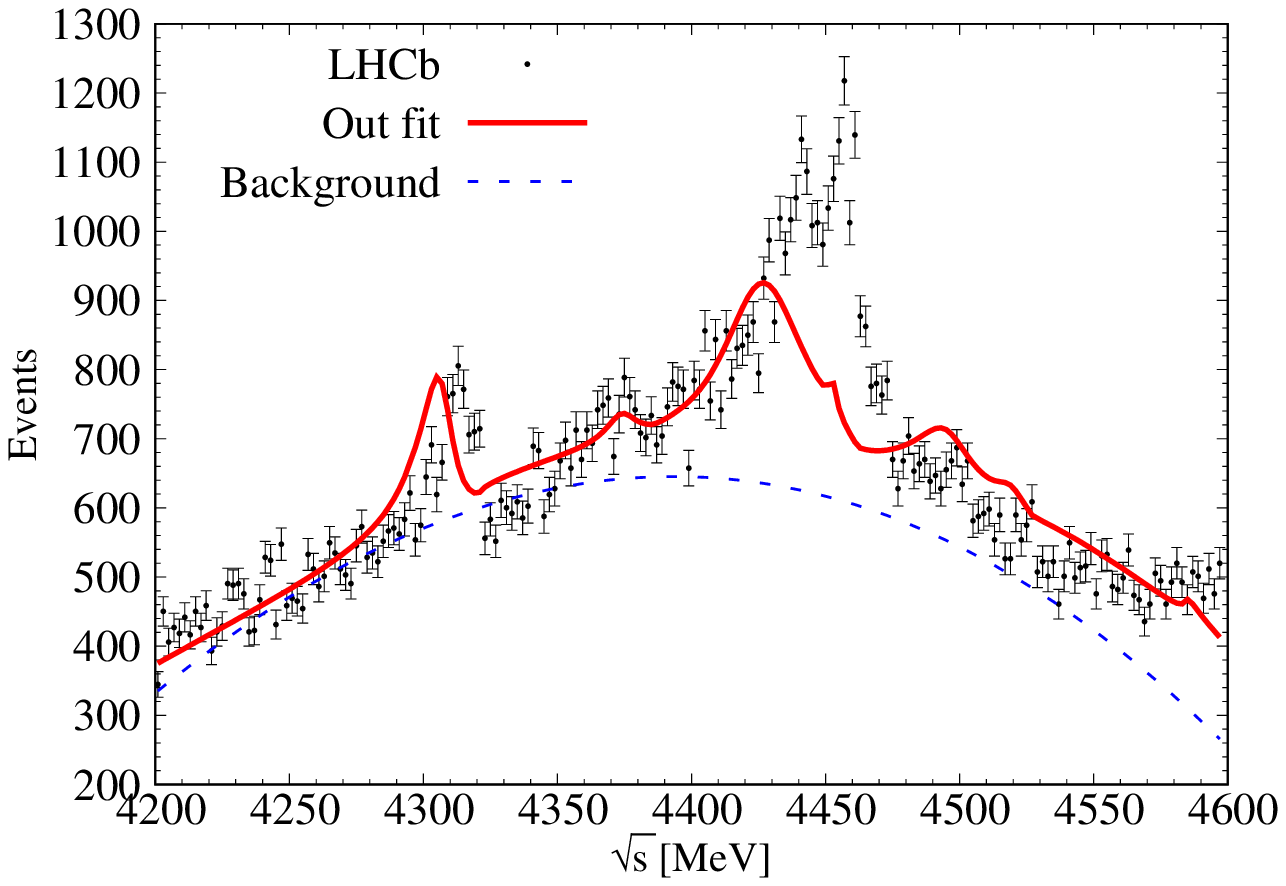}
\caption{Invariant mass distributions of $m_{J/\psi p}$ in the $\Lambda_b^0 \to J/\psi K^- p$ decay, fitted with Eqs. \eqref{eq:t12}, \eqref{eq:t32}, and using the background of Eq. \eqref{eq:bg} (left) and Eq. \eqref{eq:bg2} (right).}
\label{fig:fit}
\end{figure}

First, we assume that the $J/\psi p$ final state can be directly produced in the $\Lambda_b^0$ decay, as shown in Fig. \ref{fig:lambdec} (a), and then, these $P_c$ resonances grow up in the final state interactions, as exhibited in Fig. \ref{fig:lambdec} (b). 
Since we have assigned the three $P_c$ states with spin parity as $\frac{1}{2}^-$ and $\frac{3}{2}^-$ \cite{Xiao:2019aya}, following Ref. \cite{Xiao:2016ogq}, the $J/\psi p$ invariant mass distribution in the $\Lambda_b^0 \to J/\psi K^- p$ decay is given by,
\begin{equation}
\frac{d \Gamma (M_{inv})}{d M_{inv}} = \frac{1}{4 (2\pi)^3} \; \frac{1}{M_{\Lambda_b}} \;\tilde{q}_{J/\psi}\; q_K ( | T_{J/\psi p}^{J^P=\frac{1}{2}^-} |^2 + | T_{J/\psi p}^{J^P=\frac{3}{2}^-} |^2 ) \; ,
\end{equation}
where $M_{inv}$ is the invariant mass of the $J/\psi p$ system, the CM momenta are given by
\begin{align}
\tilde{q}_{J/\psi} (M_{inv}) &= \frac{\lambda^{1/2} (M_{inv}^2,m_{J/\psi}^2,M_p^2)}{2M_{inv}} \; ,  \\
q_K (M_{inv}) &= \frac{\lambda^{1/2} (M_{\Lambda_b}^2,m_K^2,M_{inv}^2)}{2M_{\Lambda_b}} \; ,
\end{align}
with $\lambda(a,b,c)$ the usual K\"all\'en function given in the last section, and the transition amplitudes are given by
\begin{align}
T_{J/\psi p}^{J^P=\frac{1}{2}^-} (M_{inv}) &= C^{\frac{1}{2}^-}\; G_{J/\psi p}(M_{inv}^2)\; t_{J/\psi p \to J/\psi p}(M_{inv}),   \label{eq:t12} \\
T_{J/\psi p}^{J^P=\frac{3}{2}^-} (M_{inv}) &= C^{\frac{3}{2}^-}\; G_{J/\psi p}(M_{inv}^2)\; t_{J/\psi p \to J/\psi p}(M_{inv})  \; p_K,   \label{eq:t32}
\end{align}
with $G_{J/\psi p}(M_{inv}^2)$ the loop function, and $C^{\frac{1}{2}^-}$, $C^{\frac{3}{2}^-}$ the constants which collect the CKM matrix elements and the kinematic prefactors \cite{Lu:2016roh,Roca:2016tdh} and also contain the free parameters in the fits, and we take the amplitude  $t_{J/\psi p \to J/\psi p}(M_{inv}) = T_{J/\psi N}(M_{inv})$, which is evaluated with Eq. \eqref{eq:BS} and in the isospin basis ($I=\frac{1}{2}$). 
Note that, for the amplitude $T_{J/\psi p}^{J^P=\frac{3}{2}^-} (M_{inv})$ of Eq. \eqref{eq:t32}, we have introduced an extra momentum factor for the kaon in p-wave as done in Refs. \cite{Wu:2009tu,Wu:2009nw}, and absorbed the term for the tree level contribution from the direct decay diagram~\cite{Xiao:2016ogq} of Fig. \ref{fig:lambdec} (a) into the background below.
To fit the $J/\psi p$ invariant mass distribution, we need to consider the background and we first try a low-order polynomial as suggested in Ref.  \cite{Aaij:2019vzc},
\begin{equation}
B_g = a + b s + c s^2,  \label{eq:bg}
\end{equation}
where $s$ is the Mandelstam variable of the two-body system ($s=M_{inv}^2$) and $a$, $b$, $c$ are the free parameters. 
We fit the experimental data using MINUIT \cite{James:1975dr}. 
To see the dynamical generation of the three $P_c$ states in our formalism, we first take the same value for the subtraction constant $a_\mu (\mu=1\gev) = -2.09$ in the loop function as in Ref. \cite{Xiao:2019aya}, see Eq. \eqref{eq:Gdr}, and $q_{max} = 800 \mev$ in the box diagram corrections, see Eqs. \eqref{eq:box1}-\eqref{eq:box3}, which is  the central value used in Ref. \cite{Uchino:2015uha}. 
We will come back to the choice of these values later. 
Therefore, the free parameters are the ones in Eqs. \eqref{eq:t12}, \eqref{eq:t32} and \eqref{eq:bg}. 
Our fit results are given in the left panel of Fig. \ref{fig:fit}, where one can see that the fit in the region above 4440 MeV is bad: the peak for the $P_c(4440)$  moves to lower energy, and, more unsatisfactorily the one for the $P_c(4457)$ state is nearly invisible. 
In addition, a resonance structure shows up around 4380 MeV. 
To understand the poor fit, we choose another background as used in Ref. \cite{Du:2019pij}, which is also adopted in Ref.  \cite{Aaij:2019vzc},
\begin{equation}
B_g = a + b s + c s^2 + \left| \frac{g_r}{m^2-s-i \Gamma\sqrt{s}} \right|,  \label{eq:bg2}
\end{equation}
where  $a,\ b,\ c,\ g_r,\ m,\ \Gamma$ are all free parameters for the fits. Using this one and Eqs. \eqref{eq:t12} and \eqref{eq:t32}, the fit results are shown on the right panel of Fig. \ref{fig:fit}. 
Clearly the use of a different background did not improve the description of the
data and the resulting fit is similar to the original one.

\begin{figure}
\centering
\includegraphics[width=0.48\linewidth]{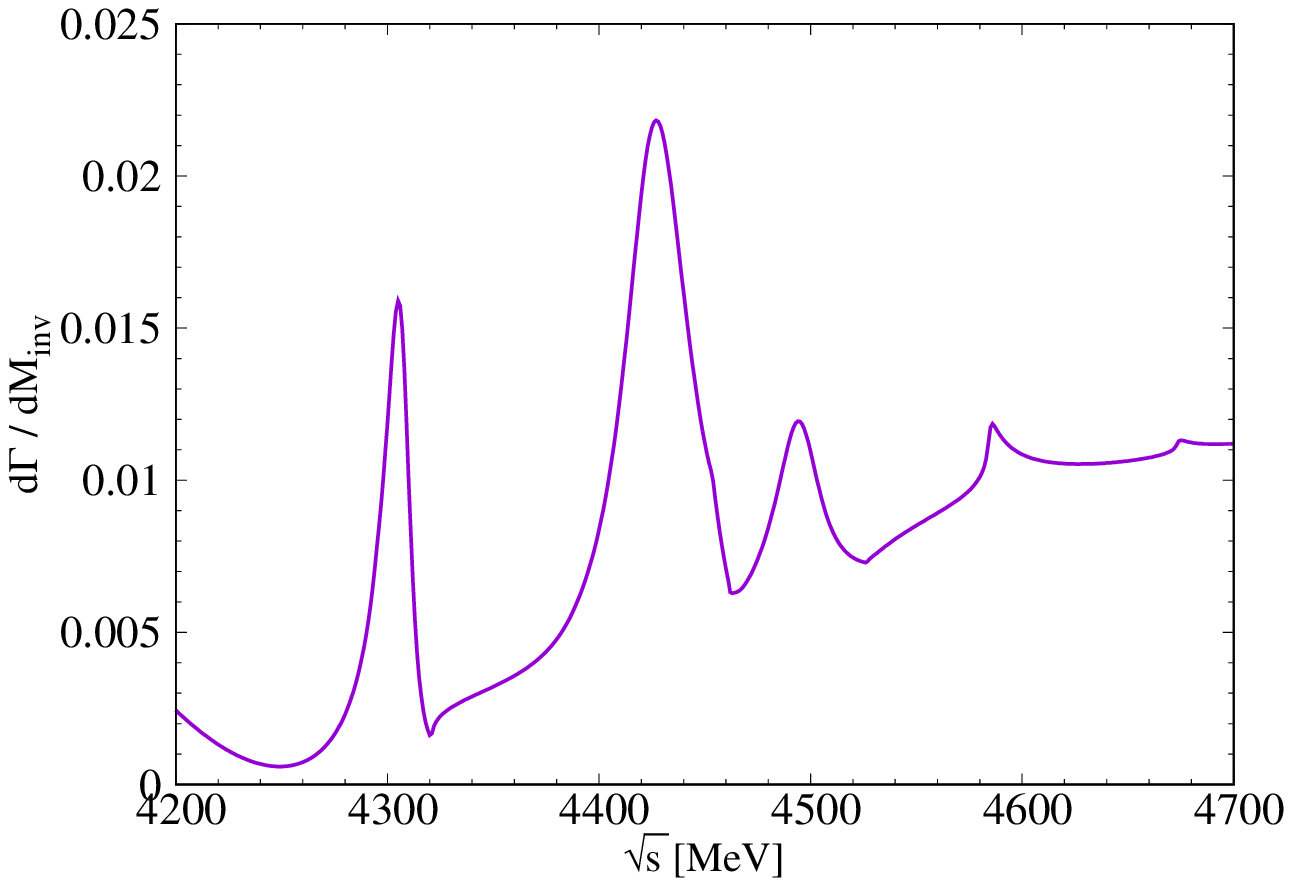}
\includegraphics[width=0.48\linewidth]{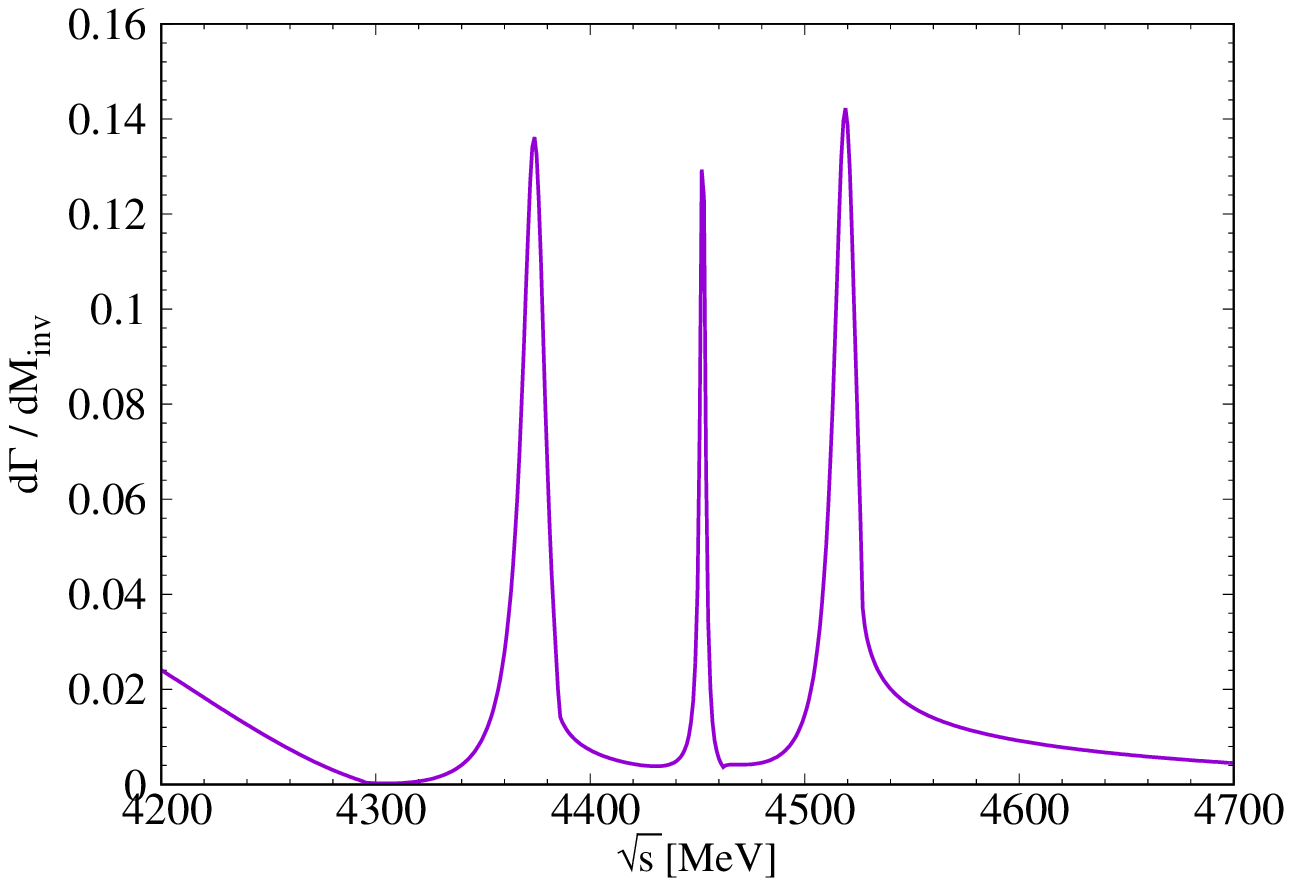}
\caption{Invariant mass distributions of $m_{J/\psi p}$ from the contribution of $T_{J/\psi p}^{J^P=\frac{1}{2}^-}$ (left) and $T_{J/\psi p}^{J^P=\frac{3}{2}^-}$ (right), respectively.}
\label{fig:masd}
\end{figure}

\begin{figure}
\centering
\includegraphics[scale=0.8]{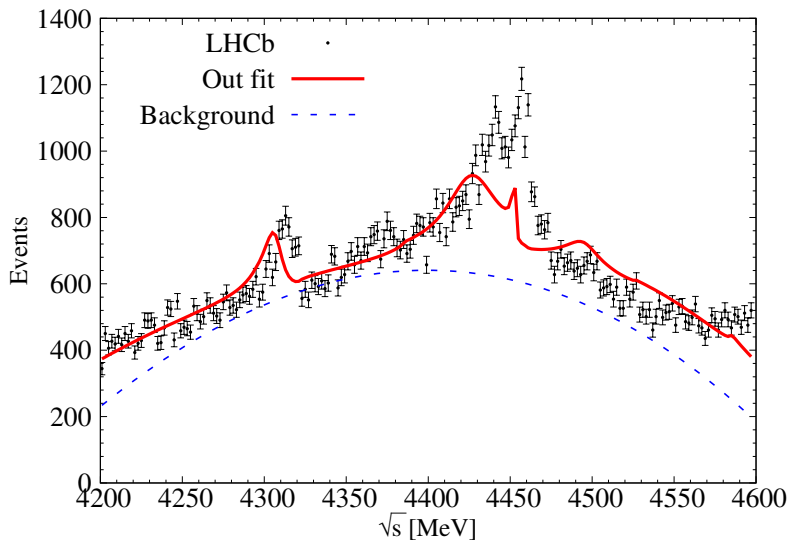}
\caption{Invariant mass distributions of $m_{J/\psi p}$ for the $\Lambda_b^0 \to J/\psi K^- p$ decay, fitted with Eqs. \eqref{eq:t12}, \eqref{eq:t32} and \eqref{eq:bg}, but only  with contribution from the 4450 pole in the $J^P = \frac{3}{2}^-$ channel.}
\label{fig:fit2}
\end{figure}

Even though we have dynamically generated three $P_c$ states in the coupled channel interactions, the $P_c(4457)$ state is almost invisible in the fit of the invariant mass distribution of $J/\psi p$. 
Note that we did not include any box diagram contributions in the $J^P = \frac{3}{2}^-$ sector. 
To understand what happened, we plot in Fig. \ref{fig:masd} the contributions of Eqs. \eqref{eq:t12} and \eqref{eq:t32} to the $J/\psi p$ invariant mass distribution separately by taking $C^{\frac{1}{2}^-} = C^{\frac{3}{2}^-} = 1$, where we can see that the contributions from the amplitude $T_{J/\psi p}^{J^P=\frac{3}{2}^-}$ to the three peaks in the $J^P = \frac{3}{2}^-$ sector are of the same magnitude. 
Therefore, the fits for the bump around the region of 4380 MeV suppress the total $T_{J/\psi p}^{J^P=\frac{3}{2}^-}$, and correspondingly, the peak of the $P_c(4457)$ disappeared with the competition of the strong one $P_c(4440)$ nearby. 
Since there are no other state (except for $P_c(4312)$, $P_c(4440)$, and $P_c(4457)$ ) claimed in the experimental invariant mass distribution $m_{J/\psi p}$~\cite{Aaij:2019vzc}, as a test, we could remove the other two contributions in the region of 4380 MeV and 4520 MeV in the $J^P = \frac{3}{2}^-$ sector. 
The corresponding fit results are shown in Fig. \ref{fig:fit2}. 
However, even though we only take one pole contribution in the $J^P = \frac{3}{2}^-$ sector, the fits in the $P_c(4457)$ region are still not good.
It is the fact that the strength of $J^P = \frac{1}{2}^-$ sector and that of $J^P = \frac{3}{2}^-$ sector are correlated, and the peak around 4312 MeV constrains the over all strength, which suppresses the fit around the 4450 MeV too.

In order to solve the problem, we have performed more fits with different options. 
First, we have checked that adding more higher order contributions to the background of Eq. \eqref{eq:bg} do not improve the fits, as commented in Ref. \cite{Aaij:2019vzc}. 
Second, for the amplitude $T_{J/\psi p}^{J^P=\frac{1}{2}^-} (M_{inv})$ of Eq. \eqref{eq:t12}, the kaon can also be in $P$-wave as discussed in Ref. \cite{Roca:2016tdh}. 
But, we find that this is not very helpful to improve the fit. 
Third, we note that it does not matter whether the tree level contribution is factored out or not, because it has been absorbed into the background as discussed above. 
Thus, we are forced to conclude that, fitted with $J/\psi p$ produced directly in the decay process, the results could not be much better than the one shown in Fig. \ref{fig:fit}, implying that the fits could not describe the experimental data well. 
Therefore, we try to improve our fit results by considering the $J/\psi p$ produced indirectly in the next section.

\section{Results with $J/\psi p$ produced indirectly}

\begin{figure}
\centering
\includegraphics[scale=0.8]{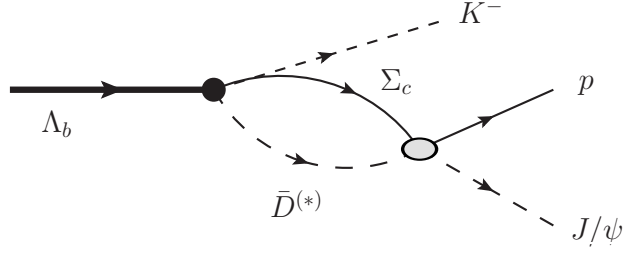}
\caption{Diagrams for the $\Lambda_b^0 \to J/\psi K^-\, p$ decay with indirect $J/\psi p$ production from the final state interactions.}
\label{fig:lambdec2}
\end{figure}

\begin{figure}
\centering
\includegraphics[width=0.48\linewidth]{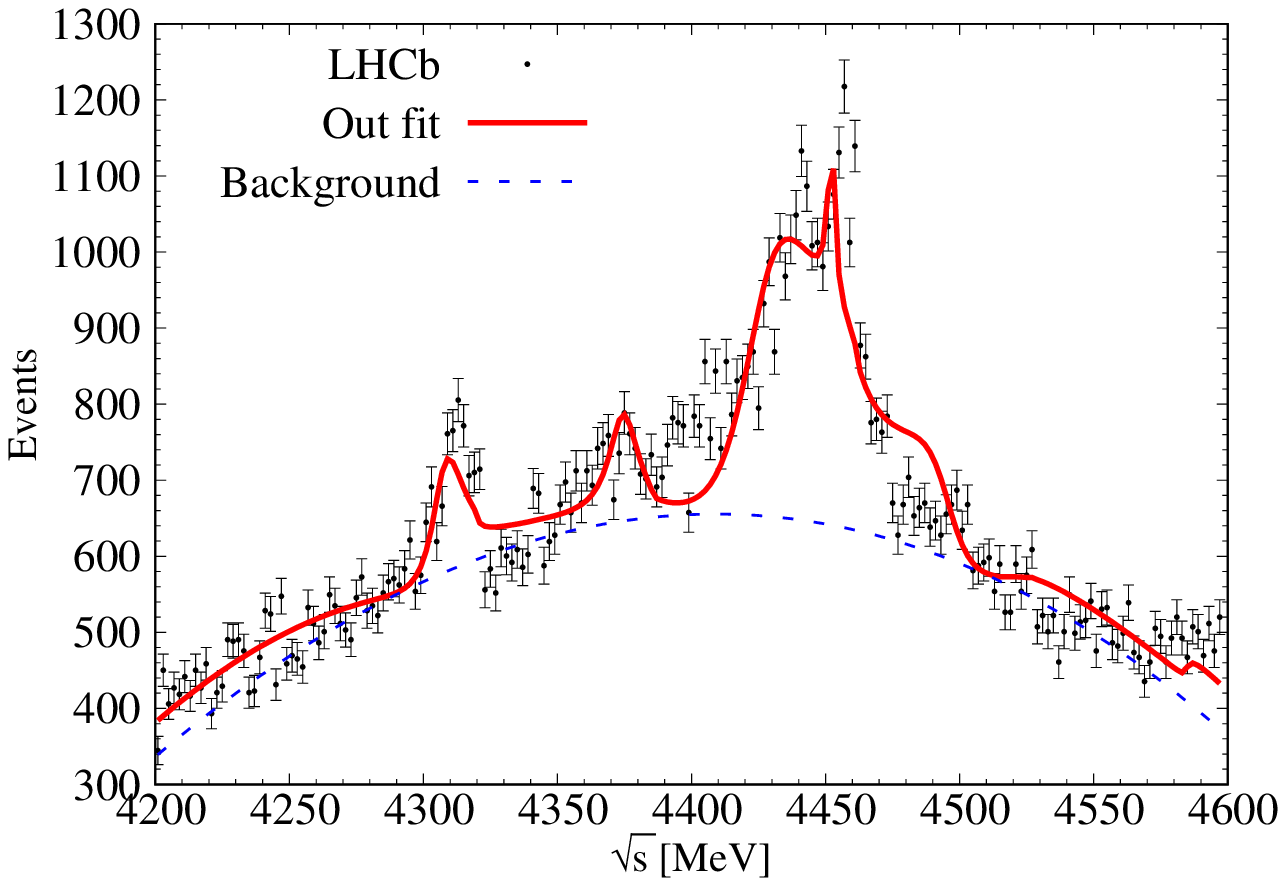}
\includegraphics[width=0.48\linewidth]{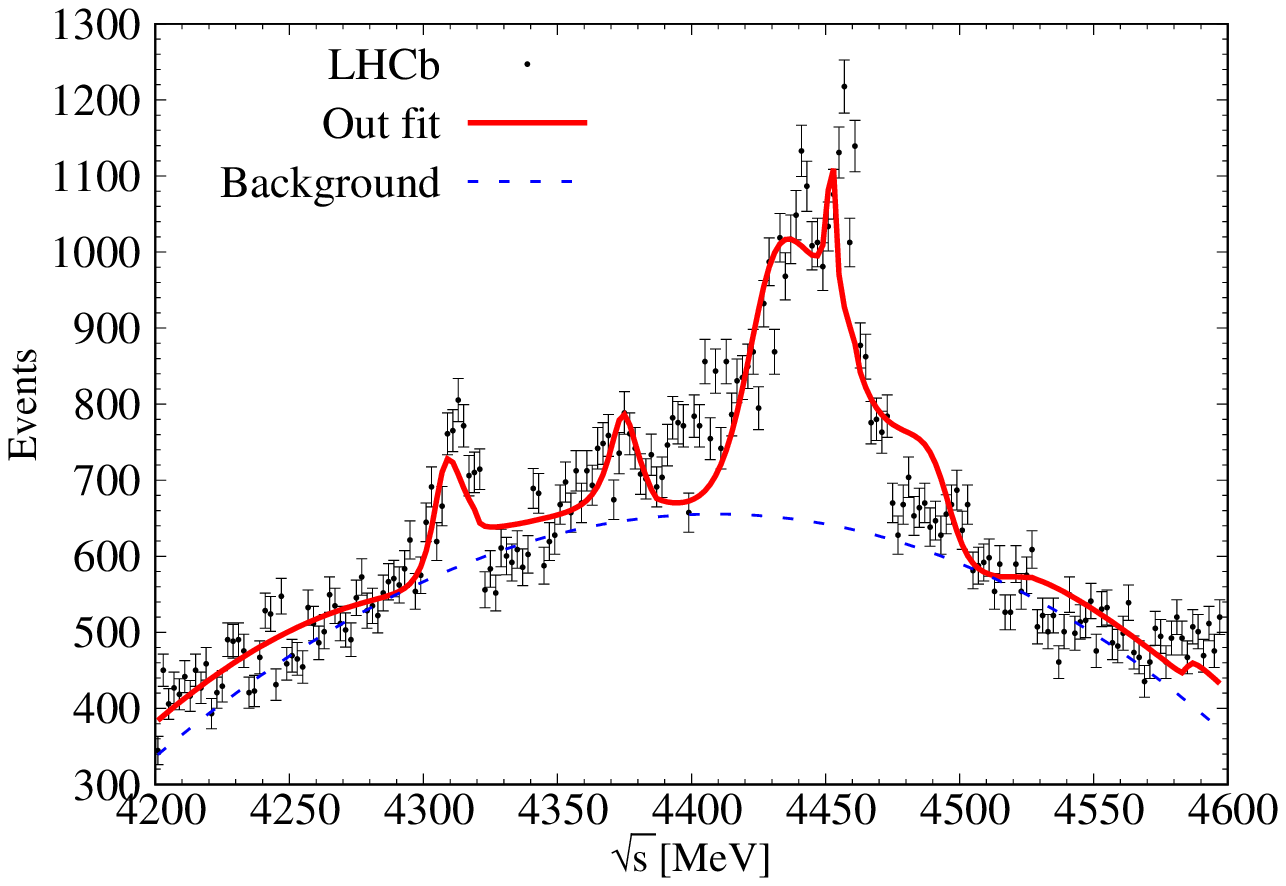}
\caption{Invariant mass distributions of $m_{J/\psi p}$ for the $\Lambda_b^0 \to J/\psi K^- p$ decay, fitted with Eqs. \eqref{eq:t12new}, \eqref{eq:t32new}, and with the background of Eq. \eqref{eq:bg} (left) and Eq. \eqref{eq:bg2} (right).}
\label{fig:fit4}
\end{figure}

\begin{figure}
\centering
\includegraphics[scale=0.8]{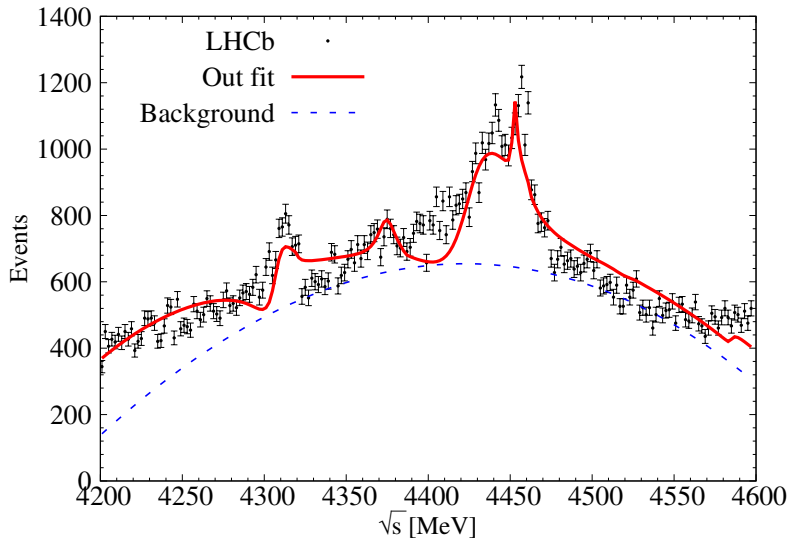}
\caption{Invariant mass distributions of $m_{J/\psi p}$ for the $\Lambda_b^0 \to J/\psi K^- p$ decay, fitted with Eqs. \eqref{eq:t12new}, \eqref{eq:t32new} \eqref{eq:bg2}, and ignoring the contributions from the $\bar{D}^*\Sigma_c^*$ channel.}
\label{fig:fit5}
\end{figure}

\begin{figure}
\centering
\includegraphics[scale=0.8]{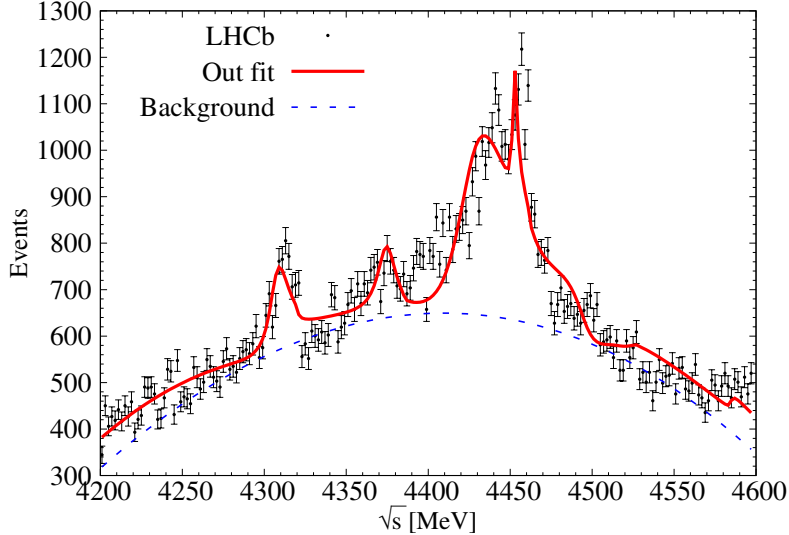}
\caption{Invariant mass distributions of $m_{J/\psi p}$ for the $\Lambda_b^0 \to J/\psi K^- p$ decay, fitted with Eqs. \eqref{eq:bg2}, \eqref{eq:t12new}, \eqref{eq:t32new}, and complex $C^{\frac{1}{2}^-}_i,\ C^{\frac{3}{2}^-}_j,\ (i,\ j=2,3)$.}
\label{fig:fit8}
\end{figure}

In the former section, we could not obtain reasonable fits with only direct $J/\psi p$ production in the $\Lambda_b^0 \to J/\psi K^-\, p$ decay. 
Indeed, the direct $J/\psi p$ production is Okubo-Zweig-Iizuka suppressed as discussed in Ref. \cite{Du:2019pij}. 
Thus, to improve the fits, we consider the indirect production process of $J/\psi p$ from final state interactions, as shown in Fig. \ref{fig:lambdec2}. 
For this decay process, we need to replace Eqs. \eqref{eq:t12} and \eqref{eq:t32} with
\begin{align}
T_{J/\psi p}^{J^P=\frac{1}{2}^-} (M_{inv}) &= C^{\frac{1}{2}^-}_1\; G_{\bar{D}\Sigma_c}(M_{inv}^2)\; t_{\bar{D}\Sigma_c \to J/\psi N}(M_{inv}) \nonumber \\
&+ C^{\frac{1}{2}^-}_2\; G_{\bar{D}^*\Sigma_c}(M_{inv}^2)\; t_{\bar{D}^*\Sigma_c \to J/\psi N}(M_{inv}) \nonumber \\
&+ C^{\frac{1}{2}^-}_3\; G_{\bar{D}^*\Sigma_c^*}(M_{inv}^2)\; t_{\bar{D}^*\Sigma_c^* \to J/\psi N}(M_{inv}) \,
,\label{eq:t12new}  \\
T_{J/\psi p}^{J^P=\frac{3}{2}^-} (M_{inv}) &= \big[ C^{\frac{3}{2}^-}_1\; G_{\bar{D}^*\Sigma_c}(M_{inv}^2)\; t_{Jbar{D}^*\Sigma_c \to J/\psi N}(M_{inv}) \nonumber \\
&+ C^{\frac{3}{2}^-}_2\; G_{\bar{D}\Sigma_c^*}(M_{inv}^2)\; t_{\bar{D}\Sigma_c^* \to J/\psi N}(M_{inv}) \nonumber \\
&+ C^{\frac{3}{2}^-}_3\; G_{\bar{D}^*\Sigma_c^*}(M_{inv}^2)\; t_{\bar{D}^*\Sigma_c^* \to J/\psi N}(M_{inv})
\big] \; p_K,  \label{eq:t32new}
\end{align}
where $C^{\frac{1}{2}^-}_i,\ C^{\frac{3}{2}^-}_j,\ (i,\ j=1,2,3)$ are free parameters. 
Our fit results are shown in Fig. \ref{fig:fit4}, where the one on the left is fitted with the background of Eq. \eqref{eq:bg} and the one on the right with Eq. \eqref{eq:bg2}. 
From Fig. \ref{fig:fit4}, one can see that the fit on the left panel for the energy range of the $P_c(4457)$ is similar to the right one, where the total $\chi^2$ and the backgrounds are not much different.
It is clear that the form of background does not influence the fit a lot.
To check the influence of the data in the energy range of 4520 MeV, we neglect the contributions of the two $\bar{D}^*\Sigma_c^*$ states by removing the terms of $C^{\frac{1}{2}^-}_3,\ C^{\frac{3}{2}^-}_3$ in Eqs. \eqref{eq:t12new} and \eqref{eq:t32new}, and obtain the results of Fig. \ref{fig:fit5}, where the fit results are not much different and only the peak of the $P_c(4440)$ is a bit lower.
It suggests that the predicted resonance around 4520 MeV may exist but more data are needed to draw a conclusion.
On the other hand, when the $C^{\frac{1}{2}^-}_i,\ C^{\frac{3}{2}^-}_j,\ (i,\ j=2,3)$ in Eqs. \eqref{eq:t12new} and \eqref{eq:t32new} are taken to be complex numbers, we obtain the results of Fig. \ref{fig:fit8}, which are similar to those of Fig. \ref{fig:fit4} and only the fit around the $P_c(4457)$ peak becomes a bit better, with a slightly smaller $\chi^2$. 
Thus, the results of Fig. \ref{fig:fit8} imply that it is not necessary to treat $C^{\frac{1}{2}^-}_i$ and $C^{\frac{3}{2}^-}_j$ as complex numbers.

\begin{figure}
\centering
\includegraphics[width=0.48\linewidth]{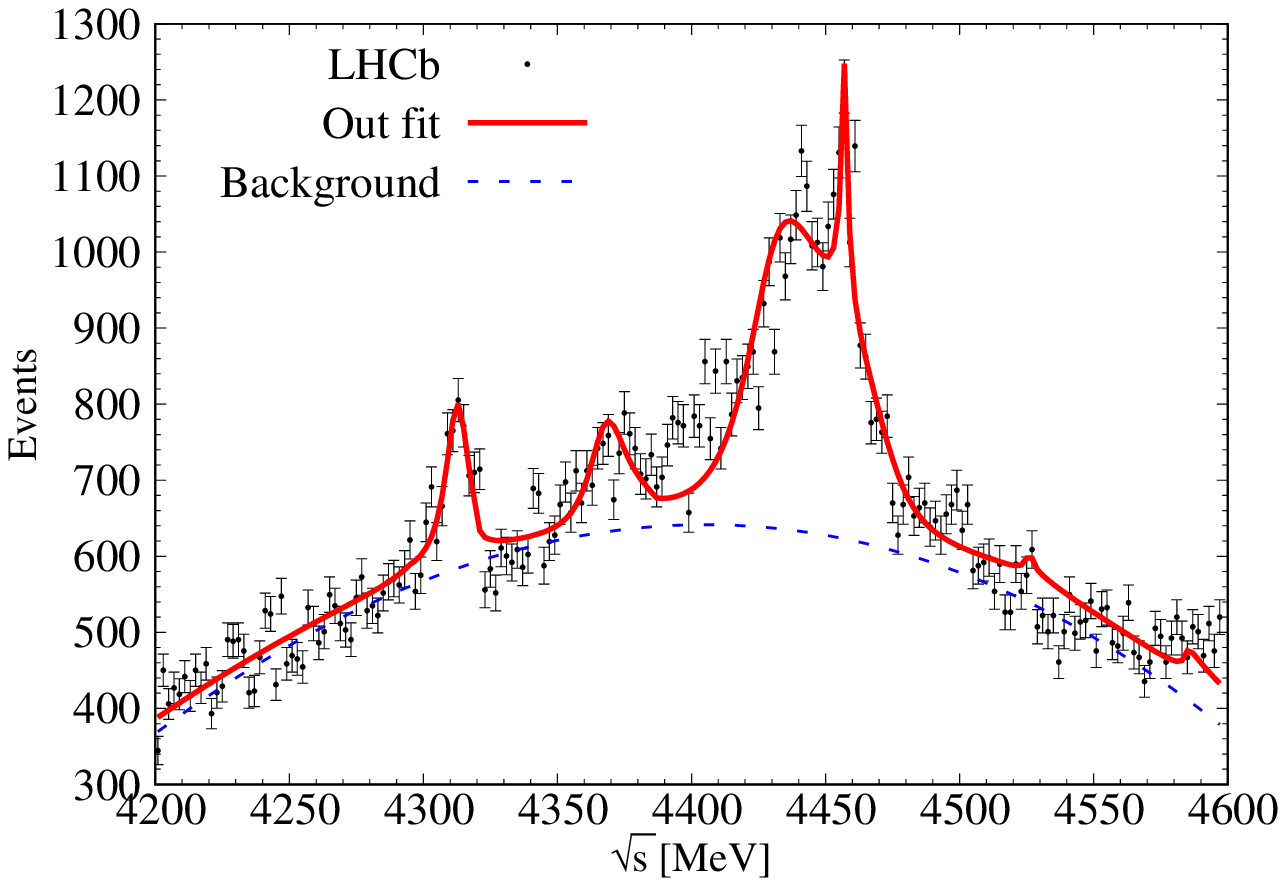}
\includegraphics[width=0.48\linewidth]{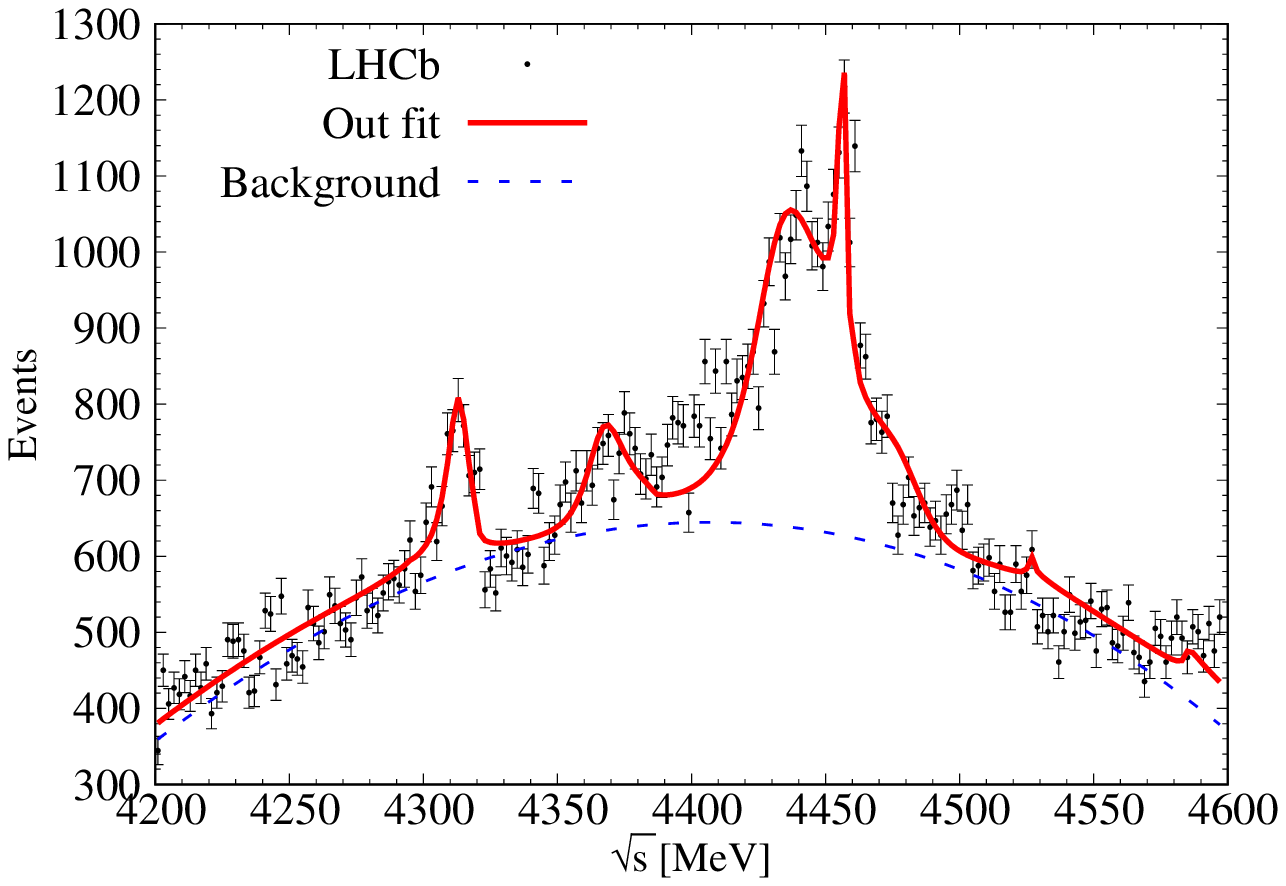}
\caption{Invariant mass distributions of $m_{J/\psi p}$ for the $\Lambda_b^0 \to J/\psi K^- p$ decay, fitted with Eqs. \eqref{eq:t12new}, \eqref{eq:t32new}, and using the background of Eq. \eqref{eq:bg} (left) and Eq. \eqref{eq:bg2} (right), where we allow $a_\mu$ for the six channels to float in the fits.}
\label{fig:fit14}
\end{figure}

One should keep in mind that, in all the fit results above, the two parameters, $a_\mu=-2.09$ \cite{Xiao:2019aya} in the loop functions and $q_{max}=800 \mev$ \cite{Uchino:2015uha} in the box diagram calculations, have been fixed as discussed after Eq. \eqref{eq:bg}. 
To obtain even better fits, we made two more attempts. 
Firstly, based on the results of Fig. \ref{fig:fit4}, we have tried to improve the fits by fixing all the other parameters except for the one of $a_\mu$. 
But, the fits are not much improved, where $a_\mu$ changes less than 0.01. 
Secondly, since the parameter $a_\mu$ can be different for each channel in principle, we allow $a_\mu$ to float for the six channels of $\bar{D}^{(*)} \Sigma_c^{(*)}$ in the fits under some constraints. 
Indeed, we obtained better results in describing the resonance structures, as shown in Fig. \ref{fig:fit14}, as expected because we introduced more free parameters 
\footnote{Note that, the results of Fig. \ref{fig:fit14}, especially the one on the right with more parameters, are not the best fits with minimum $\chi^2$, because there are much freedom with a lot of free parameters and also some fluctuations in the experimental data. 
Furthermore, floating all the free parameters with no constraint, even with only one free $a_\mu$ for all the channels, one can not obtain reasonable fits due to too much freedom for the parameters, which are not much correlated with each other, especially the one of $q_{max}=800 \mev$ in the box diagram contributions.}. 
Once again, the results with two different backgrounds do not yield visible differences in the fits (compare the left and the right panels of Fig. \ref{fig:fit14}). We would like to mention that for the fit results of Fig. \ref{fig:fit14}, the different $a_\mu$ are all within $\pm 0.06$ from the one obtained in Ref. \cite{Xiao:2019aya}, $a_\mu=-2.09$, and thus, these differences can be treated as our theoretical uncertainties. 
Besides, from the results of Fig. \ref{fig:fit14}, the $a_\mu$ for the two $\bar{D}^{*} \Sigma_c^{*}$ channels have large uncertainties because of the data fluctuated around the 4520 MeV region, as indicted in Fig. \ref{fig:fit5}.
Indeed, the fits of Fig. \ref{fig:fit14} with more free $a_\mu$ are just a bit better in the peak regions than the ones of Fig. \ref{fig:fit4}, where one can see that our theoretical model is powerful with quite a few parameters to well describe the experimental data. 
Thus, these results also confirmed the ones obtained with $a_\mu=-2.09$ in Ref. \cite{Xiao:2019aya}. 
Therefore, we choose our main results as those from the dynamical reproduction of the three $P_c$ states with only two parameters, $a_\mu=-2.09$ and $q_{max}=800 \mev$. It should be stressed that from these fit results, there seems to be a clear indication of a narrow $P_c(4380)$ apart from the three $P_c$ states reported by the LHCb Collaboration, which is also found in Ref.  \cite{Du:2019pij}, and there are no clear signals for the two $\bar{D}^* \Sigma_c^*$ states around the region of 4520 MeV in the $J/\psi p$ invariant mass distributions.

\section{Couplings, partial decay widths and predicted invariant mass distributions}

From the results in the last section, one can see that our theoretical model can describe the $m_{J/\psi p}$ invariant mass distributions reasonably well with only two parameters, $a_\mu=-2.09$ and $q_{max}=800 \mev$, in the theoretical model, of course associated with  some parameters for the background. 
Based on these results, we calculate the couplings, the partial decay widths, and the branching ratios of these $P_c$ states in the sectors of $J^P=\frac{1}{2}^-$ and $J^P=\frac{3}{2}^-$, and the results are given in Tables \ref{tab:coup11} and \ref{tab:coup31}, respectively. 
Since we have added the box diagram contributions to the channels of $\bar{D}^*\Sigma_c$ and $\bar{D}^*\Sigma_c^*$ in the $J^P=\frac{1}{2}^-$ sector, the poles and the couplings in this sector are evaluated with the method described in Ref. \cite{Uchino:2015uha}, where the position of the pole $M_p$ and the width $\Gamma_p$ are taken from the peak of the square of the scattering amplitude $T_{dd}(\sqrt{s})$ as usually done in experimental analyses, and the couplings to different channels are given by
\begin{equation}
g_d= \sqrt{\left| \frac{\Gamma_p}{2} \text{Im} T_{dd} (M_p) \right|} \, , \qquad
g_i= \frac{\text{Im} T_{id}(M_p) }{\text{Im} T_{dd}(M_p) } g_d,
\end{equation}
where the index $d$ refers to the dominant channel and the index $i$ other coupled channels.
 
From Table \ref{tab:coup11}, one can see that the pole at $(4306.0+i7.0)$ MeV is dominated by the $\bar{D} \Sigma_c$ channel and therefore assigned to the $P_c(4312)$, whereas the one of $(4433.0+i11.0)$ MeV strongly couples to the $\bar{D}^* \Sigma_c$ channel and therefore assigned to the $P_c(4440)$. 
In Table \ref{tab:coup31}, the pole at $(4452.5+i1.5)$ MeV, to which the main channel contributed is the $\bar{D}^* \Sigma_c$ channel, is assumed to be the $P_c(4457)$ for its small width. 
There are two other states around 4500 MeV and 4520 MeV, respectively, which are dominated by the $\bar{D}^* \Sigma_c^*$ channel \footnote{In fact, there is another pole with $J^P = \frac{5}{2}^-$, which is also dominated by the $\bar{D}^* \Sigma_c^*$ channel but not discussed here.} and not degenerated now compared to the ones in Ref.  \cite{Xiao:2019aya} due to the box diagram contributions from the pion exchange introduced in the $J^P = \frac{1}{2}^-$ sector. 
The partial decay widths given in Table \ref{tab:coup11} show that the $P_c(4312)$ , a $\bar{D} \Sigma_c$ bound state, has a large decay width into $\eta_c N$, and a not so small one to $J/\psi N$, similar to the bound state of $\bar{D}^* \Sigma_c^*$ around 4500 MeV. 
By contrast, the $P_c(4440)$,  a $\bar{D}^* \Sigma_c$ bound state, decays mostly to $J/\psi N$ but not so much to $\eta_c N$. 
One more thing to be noted is that all of the three bound states in the $J^P = \frac{1}{2}^-$ sector have quite small partial decay widths into $\bar{D} \Lambda_c$ and $\bar{D}^* \Lambda_c$, and thus, they can not be easily observed in these two decay channels.
See also the invariant mass distributions to be discussed later. 
The reason is that in our model now we have not included off-diagonal interaction between $\bar{D}^{(*)} \Lambda_c$ and $\bar{D}^{(*)} \Sigma_c$ with the pion exchange as discussed before, and we need some improvement in the future. 
By contrast, some other work suggest $\bar{D}^{(*)} \Lambda_c$ channels may be the main decay channels of $P_c$ states.
In Refs.~\cite{Shen:2016tzq,Lin:2017mtz,Lin:2019qiv} the large decay width of $\bar{D}^{(*)} \Lambda_c$ was predicted through triangle loop diagrams by single $\pi$ exchange.
Ref.~\cite{Yamaguchi:2016ote} considered the interactions between $\bar{D}^{(*)} \Lambda_c$ and $\bar{D}^{(*)} \Sigma_c$, but the predicted mass of resonances are too low, around 4136 MeV. 
In the framework of an extended chromomagnetic model \cite{Weng:2019ynv}, the three $P_c$ states and also the predicted $P_c(4380)$ state are predicted to decay dominantly to $\bar{D}^*\Lambda_c$.
Whereas, now our calculations show that $\bar{D}^{(*)} \Lambda_c$ may exhibit weak signal of $P_c$ states. 
On the other hand, the three resonances in the $J^P = \frac{3}{2}^-$ sector, the predicted $P_c(4380)$  of $\bar{D} \Sigma_c^*$, the $P_c(4457)$  of $\bar{D}^* \Sigma_c$  and another predicted $P_c(4520)$  of $\bar{D}^* \Sigma_c^*$, decay mainly into $J/\psi N$, and negligibly into other channels. 
One should note that there are some theoretical uncertainties for the results of the partial decay widths and branching ratios in Tables  \ref{tab:coup11} and  \ref{tab:coup31}, since they are evaluated with the couplings obtained. 
Once again, we summarize the results for the three $P_c$ states reported by the LHCb Collaboration in Table \ref{tab:sum}.

\begin{table}[ht]
     \renewcommand{\arraystretch}{1.2}
     \setlength{\tabcolsep}{0.4cm}
\centering
\caption{Dimensionless coupling constants of the $(I=1/2, J^P=1/2^-)$ poles found in this
  work to different coupled channels. The imaginary part of the energies corresponds to $\Gamma/2$.} \label{tab:coup11}
\begin{tabular}{cccc cccc}
\hline\hline
$(4306.0+i7.0)$ MeV   & $\eta_c N$ & $J/\psi N$ & $\bar{D} \Lambda_c$ & $\bar{D} \Sigma_c$ & $\bar{D}^* \Lambda_c$ & $\bar{D}^* \Sigma_c$ & $\bar{D}^* \Sigma_c^*$  \\
\hline
$|g_i|$ & $0.59$ & $0.41$ & $0.01$ & $\mathbf{1.99}$ & $0.10$ & $0.02$ & $0.03$  \\
$\Gamma_i$ & $\mathbf{9.7}$ & $3.9$ & $0.0$ & -- & $0.1$ & -- & --  \\
$Br$ & $\mathbf{69.0\%}$ & $27.6\%$ & $0.0\%$ & -- & $0.9\%$ & -- & --  \\
\hline
$(4433.0+i11.0)$ MeV    & $\eta_c N$ & $J/\psi N$ & $\bar{D} \Lambda_c$ & $\bar{D} \Sigma_c$ & $\bar{D}^* \Lambda_c$ & $\bar{D}^* \Sigma_c$ & $\bar{D}^* \Sigma_c^*$  \\
\hline
$|g_i|$ & $0.16$ & $0.49$ & $0.03$ & $0.07$ & $0.03$ & $\mathbf{2.42}$ & $0.06$  \\
$\Gamma_i$ & $0.7$ & $\mathbf{6.4}$ & $0.1$ & $0.2$ & $0.0$ & -- & --  \\
$Br$ & $3.4\%$ & $\mathbf{29.0\%}$ & $0.3\%$ & $1.1\%$ & $0.2\%$ & -- & --   \\
\hline
$(4500.0+i5.5)$ MeV   & $\eta_c N$ & $J/\psi N$ & $\bar{D} \Lambda_c$ & $\bar{D} \Sigma_c$ & $\bar{D}^* \Lambda_c$ & $\bar{D}^* \Sigma_c$ & $\bar{D}^* \Sigma_c^*$  \\
\hline
$|g_i|$ & $0.37$ & $0.26$ & $0.05$ & $0.03$ & $0.02$ & $0.02$ & $\mathbf{2.29}$  \\
$\Gamma_i$ & $\mathbf{4.5}$ & $1.9$ & $0.2$ & $0.1$ & $0.0$ & $0.0$ & --  \\
$Br$ & $\mathbf{41.2\%}$ & $17.7\%$ & $1.5\%$ & $0.5\%$ & $0.3\%$ & $0.0\%$ & --  \\
\hline
\end{tabular}
\end{table}

\begin{table}[ht]
     \renewcommand{\arraystretch}{1.2}
         \setlength{\tabcolsep}{0.4cm}
\centering
\caption{Same as Table~\ref{tab:coup11} but for $J^P=3/2^-$.} \label{tab:coup31}
\begin{tabular}{ccc ccc}
\hline\hline
$(4374.3+i6.9)$ MeV & $J/\psi N$ & $\bar{D}^* \Lambda_c$ & $\bar{D}^* \Sigma_c$ & $\bar{D} \Sigma_c^*$ & $\bar{D}^* \Sigma_c^*$  \\
\hline
$|g_i|$ & $0.73$ & $0.18$ & $0.19$ & $\mathbf{1.94}$ & $0.30$  \\
$\Gamma_i$ & $\mathbf{13.5}$ & $1.1$ & -- & -- & --  \\
$Br$ & $\mathbf{98.4\%}$ & $7.7\%$ & -- & -- & --  \\
\hline
$(4452.5+i1.5)$ MeV & $J/\psi N$ & $\bar{D}^* \Lambda_c$ & $\bar{D}^* \Sigma_c$ & $\bar{D} \Sigma_c^*$ & $\bar{D}^* \Sigma_c^*$  \\
\hline
$|g_i|$ & $0.30$ & $0.07$ & $\mathbf{1.82}$ & $0.08$ & $0.19$   \\
$\Gamma_i$ & $\mathbf{2.6}$ & $0.2$ & -- & $0.2$ & --  \\
$Br$ & $\mathbf{85.9}$ & $6.9$ & -- & $8.3$ & --  \\
\hline
$(4519.0+i6.9)$ MeV & $J/\psi N$ & $\bar{D}^* \Lambda_c$ & $\bar{D}^* \Sigma_c$ & $\bar{D} \Sigma_c^*$ & $\bar{D}^* \Sigma_c^*$  \\
\hline
$|g_i|$ & $0.66$ & $0.13$ & $0.10$ & $0.13$ & $\mathbf{1.82}$   \\
$\Gamma_i$ & $\mathbf{12.7}$ & $0.9$ & $0.3$& $0.8$ & --  \\
$Br$ & $\mathbf{92.4\%}$ & $6.7\%$ & $2.5\%$& $5.8\%$ & --  \\
\hline
\end{tabular}
\end{table}

\begin{table}[htb]
\renewcommand{\arraystretch}{1.7}
     \setlength{\tabcolsep}{0.2cm}
\centering
\caption{Main results for the three $P_c$ states found experimentally~\cite{Aaij:2019vzc} (unit: MeV).}
\label{tab:sum}
\begin{tabular}{lccc cc}
\hline
Mass  & Width  & Bound channel  & $J^P$   &  Experiments (mass, width)   \\
\hline\hline
4306.4 & 14.0 & $\bar D \Sigma_c$      &  $1/2^-$ & $P_c(4312)$: (4311.9, 9.8)   \\
\hline
4433.0 & 22.0 & $\bar D^* \Sigma_c$   & $1/2^-$ &  $P_c(4440)$: (4440.3, 20.6)  \\
\hline
4452.5 & 3.0    & $\bar D^* \Sigma_c$   & $3/2^-$ &  $P_c(4457)$: (4457.3, 6.4)  \\
\hline
\end{tabular}
\end{table}

Note that our results for the partial decay widths and branching ratios in Tables~\ref{tab:coup11} and \ref{tab:coup31} are compatible with other results obtained with different theoretical models. 
Within the framework of the Bethe-Salpeter equation with the effective interactions provided by light vector meson exchanges from the chiral Lagrangian, where the dynamics is analogous to ours as depicted in Fig. \ref{fig:PVB}, a decay width of $\Gamma[P_c(4312)\to J/\psi p]=3.66\mev$ is obtained in Ref. \cite{Ke:2019bkf}, which is consistent with ours, 3.9 MeV, within uncertainties. 
Using an effective Lagrangian approach, the partial widths for $P_c(4312)$, $P_c(4440)$, $P_c(4457)$ decaying to $J/\psi p$ are found to be 5.03 MeV, 9.38 MeV, and  2.89 MeV, respectively, with a cutoff 1.0 GeV in Ref. \cite{Xiao:2019mst}, which within uncertainties are consistent with ours and these results for the decay fractions and productions are further discussed in Ref. \cite{Wu:2019rog}.
By contrast, the dominant decay channels of the $P_c(4312)$ are found to be $J/\psi p$ and $\eta_c p$, incompatible with the results of the chiral constituent quark model~\cite{Dong:2020nwk}. 
As discussed in Ref. \cite{Cheng:2019obk}, the branching ratios for the three $P_c$ states decaying to $J/\psi p$ are no more than 2\% in the molecular picture, even in the compact pentaquark picture the partial decay widths are quite small too. 
Different results for the three $P_c$ states decaying to $J/\psi p$ and $\eta_c p$ are also given in Ref. \cite{Wang:2019spc} based on the quark interchange model. More concerns about the $P_c$ states decay into  $J/\psi p$ and $\eta_c p$ can be referred to Refs. \cite{Voloshin:2019aut,Sakai:2019qph}.

\begin{figure}
\centering
\includegraphics[scale=0.8]{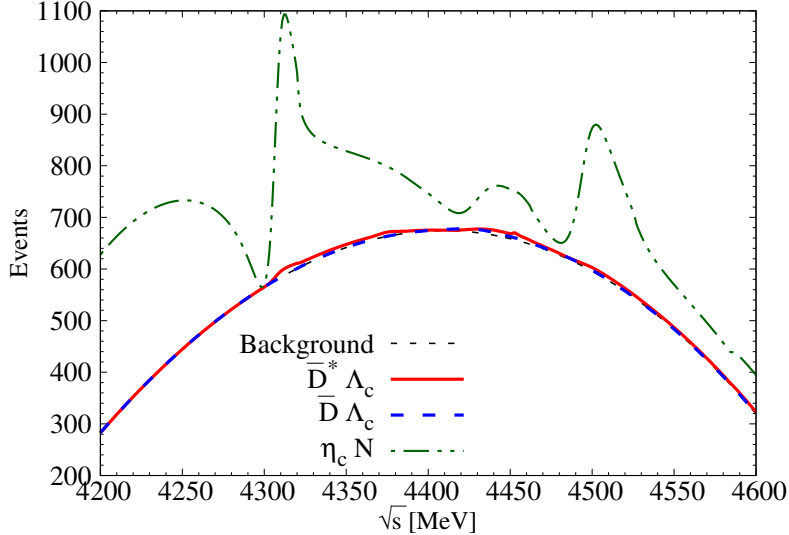}
\caption{Predicted invariant mass distributions for the channels of $\bar{D}^* \Lambda_c$, $\bar{D} \Lambda_c$ and $\eta_c N$.}
\label{fig:premas}
\end{figure}

Furthermore, to provide references for searches for these states in other decay channels, adopting the decay procedure of Fig. \ref{fig:lambdec2} and based on the fit results obtained above, we predict the invariant mass distributions in the $\bar{D}^* \Lambda_c$, $\bar{D} \Lambda_c$ and $\eta_c N$ channels by changing the final decay channels in the coupled channel interactions.  
The results are shown in  Fig. \ref{fig:premas}, where we have taken the same background of Fig. \ref{fig:fit4}. 
From Fig. \ref{fig:premas}, one can see that only the $P_c(4312)$ and the predicted $P_c(4520)$ are clearly visible in the $\eta_c N$ decay channel, whereas the $P_c(4440)$ shows up as a structure at higher energy  around 4450 MeV. 
It is also expected in Ref. \cite{Wang:2019spc} that future experiment can search for the $P_c(4312)$ state in the $\eta_c N$ channel. 
By contrast, it is obvious that all the resonances both in the $J^P = \frac{1}{2}^-$ and $J^P = \frac{3}{2}^-$ sectors can not be seen in the $\bar{D}^* \Lambda_c$ channel, and the ones in the $J^P = \frac{1}{2}^-$ sector can not be found in the $\bar{D} \Lambda_c$ channel either. 
In fact, from Tables \ref{tab:coup11} and \ref{tab:coup31}, one can see that the partial decay widths into $\bar{D}^{(*)} \Lambda_c$ are nearly zero in the $J^P=\frac{1}{2}^-$ sector and quite small in the $J^P=\frac{3}{2}^-$ sector. 
As we discussed before, it is the fact that in our model we miss the off-diagonal interaction between $\bar{D}^{(*)} \Lambda_c$ and $\bar{D}^{(*)} \Sigma_c$ with the pion exchange, after making the improvement on this point in the future, the predictions about the decay width of $\bar{D}^{(*)} \Lambda_c$ channels may be further improved.
In the current work, our main aims are to solve the mass degeneracy problem and to examine the $P_c$ states just from the pure $J/\psi p$ final interaction or the full coupled channel effect, thus, the current model is enough to clarify these problems.

\section{Conclusions}

In the present work, we revisited the interactions of ${\bar D}^{(*)}\Sigma_c^{(*)}$ and their coupled channels with the chiral unitary approach and the constraints of heavy quark spin symmetry. 
By taking into account pion exchanges in the interaction potentials of the main channels in the $J^P=\frac{1}{2}^-$ sector, which are introduced via box diagram contributions, the degeneracy in the masses of the two $\bar{D}^* \Sigma_c$ states with spin parities of $J^P=\frac{1}{2}^-$ and $J^P=\frac{3}{2}^-$ is lifted, compared with Ref. \cite{Xiao:2019aya}, which are assigned as the $P_c(4440)$ and  $P_c(4457)$, respectively, reported by the LHCb Collaboration with the updated data of Run II. 
Thus, based on this new model, we performed several fits of the $J/\psi p$ invariant mass distributions in the $\Lambda_b^0 \to J/\psi K^- p$ decay to examine the nature of the three $P_c$ states.

We first fitted with the $J/\psi p$ directly produced in the $\Lambda_b^0$ decay. 
From the fit results, we found that it is difficult to describe the experimental data reasonably well with the three $P_c$ states showing up in the invariant mass distributions, even though we tried different choices for the background and other possible methods. 
Indeed, the $J/\psi p$ direct production process in the $\Lambda_b^0$ decay is Okubo-Zweig-Iizuka suppressed \cite{Du:2019pij}. 
Thus, the $J/\psi p$ indirect production in the final state interactions of the $\Lambda_b^0$ decay products is utilized in the next fit. 
In this fitting procedure, even we only used one free parameter for the meson-baryon loop functions, the experimental data can be fitted well, where three $P_c$ states appear and one more narrow $P_c(4380)$ state is predicted, as in Refs. \cite{Liu:2019tjn,Xiao:2019aya,Du:2019pij,Yamaguchi:2019seo}. 
Of course, when we took more free parameters for the loop functions, we obtained better but qualitatively the same fit results. 
Upon examining the later fit results, it is clear that all these $P_c$ states (also the predicted one around 4380 MeV) are dynamically reproduced in the final state interactions of the $\Lambda_b^0 \to J/\psi K^- p$ decay, which provides a non-trivial confirmation of
 their molecular nature. 

With the fit results obtained, we evaluated the partial decay widths and the branching ratios of these $P_c$ states to other decay channels. 
Assuming the same background contributions as the analogous indirect productions $J/\psi p$ in the $\Lambda_b^0$ decay, we predicted the invariant mass distributions for the channels of $\bar{D}^* \Lambda_c$, $\bar{D} \Lambda_c$ and $\eta_c N$. 
Both in the results of the partial decay widths and the predicted invariant mass distributions, the $P_c(4312)$ state and the one of $\bar{D}^* \Sigma_c^*$ around 4500 MeV are shown to have large partial decay widths into $\eta_c N$  and clear resonance signals are seen in the corresponding invariant mass distributions. 
On the other hand, there are also some contributions for the partial decay widths and the invariant mass distributions from the $P_c(4440)$, a bound state of $\bar{D}^* \Sigma_c$.
Whereas, the $P_c(4457)$ (a loosely bound state of $\bar{D}^* \Sigma_c$) and the other two molecules of $\bar{D} \Sigma_c^*$, $\bar{D}^* \Sigma_c^*$ around  4380 MeV, 4520 MeV, respectively, do not decay into $\eta_c N$ due to their predicted spin parity as $J^P = \frac{3}{2}^-$. 
Therefore, it is crucial to search for these $P_c$ states in the $\eta_c N$ channel to distinguish their different structure and spin properties, especially for the two $\bar{D}^* \Sigma_c$ bound states, $P_c(4440)$ and $P_c(4457)$. 
We hope that our results can be tested by other theoretical models and future experiments.
Furthermore, we find that these $P_c$ states and the other predicted ones have very small decay width to the channels of $\bar{D}^* \Lambda_c$ and $\bar{D} \Lambda_c$, because in the current model the interaction between $\bar{D}^{(*)} \Lambda_c$ and $\bar{D}^{(*)} \Sigma_c$ through single $\pi$ exchange is still missing. 
In the future, we may extend our model to recover these interactions to make even full coupled channel interaction to reveal the nature of these $P_c$ states.
It is a challenging work because as shown in Ref.~\cite{Yamaguchi:2016ote} when the interaction between $\bar{D}^{(*)} \Lambda_c$ and $\bar{D}^{(*)} \Sigma_c$ is considered, the predicted mass of resonances are too low, around 4136 MeV.

\section*{Acknowledgments}

We thank E. Oset and J. Nieves for useful discussions and valuable comments, 
E. Oset, J. J. Xie and K.  Azizi for careful reading the manuscript and helpful suggestions, 
and acknowledge J. B. He and L. M. Zhang for useful information on the experiment.
This work is partly supported by
the National Natural Science Foundation of China under Grants Nos.11735003, 11975041, and 11961141004 (L.S.G),
and the Fundamental Research Funds for the Central Universities (J.J.W).

\end{document}